\documentclass[twocolumn, times, trackchanges, tighten]{aastex63}

\usepackage{amsmath,amstext}
\usepackage[figure,figure*]{hypcap}
\usepackage{newtxmath}

\newcommand{\biborder}[1]{}

\defcitealias{Worseck2019}{W19}

\begin{document}

\title{
New Evidence for Extended \ion{He}{2} Reionization at $z \gtrsim 3.5$ 
from \ion{He}{2} Lyman Alpha and Beta Transmission Spikes
\footnote{Based on observations made with the NASA/ESA 
\textit{Hubble Space Telescope}, obtained at the Space Telescope Science 
Institute, which is operated by the Association of Universities for Research 
in Astronomy, Inc., under NASA contract NAS5-26555. These observations are 
associated with program 15356.}
}

\received{2020 December 14}
\revised{2021 March 9}
\accepted{2021 March 10}

\shorttitle{Extended \ion{He}{2} Reionization from \ion{He}{2} Lyman Alpha and Beta Transmission Spikes}
\shortauthors{Makan et al.}

\correspondingauthor{Kirill Makan}
\email{kmakan@astro.physik.uni-potsdam.de}

\author[0000-0003-3157-1191]{Kirill Makan}
\affiliation{Institut f\"ur Physik und Astronomie, Universit\"at Potsdam, Karl-Liebknecht-Str. 24/25, 
D-14476 Potsdam, Germany}

\author[0000-0003-0960-3580]{G\'abor Worseck}
\affiliation{Institut f\"ur Physik und Astronomie, Universit\"at Potsdam, Karl-Liebknecht-Str. 24/25, 
D-14476 Potsdam, Germany}

\author[0000-0003-0821-3644]{Frederick B. Davies}
\affiliation{Department of Physics, University of California, Santa Barbara, CA 93106-9530, USA}
\affiliation{Lawrence Berkeley National Laboratory, 1 Cyclotron Rd., Berkeley, CA 94720-8139, USA}
\affiliation{Max-Planck-Institut f\"ur Astronomie, K\"onigstuhl 17, D-69117 Heidelberg, Germany}

\author[0000-0002-7054-4332]{Joseph F. Hennawi}
\affiliation{Department of Physics, University of California, Santa Barbara, CA 93106-9530, USA}

\author[0000-0002-7738-6875]{J. Xavier Prochaska}
\affiliation{University of California Observatories, Lick Observatory, University of California, Santa Cruz, CA 95064, USA}

\author[0000-0002-1188-1435]{Philipp Richter}
\affiliation{Institut f\"ur Physik und Astronomie, Universit\"at Potsdam, Karl-Liebknecht-Str. 24/25, 
	D-14476 Potsdam, Germany}

\begin{abstract}
We present new high-resolution ($R=\lambda/\Delta\lambda \sim 14,000$) 
spectra of the two brightest \ion{He}{2}-transparent quasars in the far-UV (FUV) 
at $z_{\mathrm{em}} > 3.5$, HE2QS\,J2311$-$1417 ($z_\mathrm{em} = 3.70$) 
and HE2QS\,J1630$+$0435 ($z_\mathrm{em} = 3.81$), obtained with the Cosmic 
Origins Spectrograph (COS) on the {\it Hubble Space Telescope} (\textit{HST}).
In the predominantly saturated \ion{He}{2} absorption spectra, both sightlines 
show several isolated resolved (full width at half maximum 
FWHM $\gtrsim 50\mathrm{\,km\,s^{-1}}$) transmission spikes in 
\ion{He}{2} Ly$\alpha$ and \ion{He}{2} Ly$\beta$. 
The incidence of such spikes decreases with increasing redshift, but both 
sightlines show significant spikes at $z \gtrsim 3.5$, signaling the presence 
of fully ionized regions in the $z \gtrsim 3.5$ intergalactic medium (IGM).
We employ an automated algorithm to compare the number of detected \ion{He}{2} 
transmission spikes to predictions from mock spectra created from the outputs 
of a cubic $(146\mathrm{\,cMpc})^{3}$ optically thin \texttt{Nyx} 
hydrodynamical simulation, assuming a range of UV background photoionization 
rates $\Gamma_\mathrm{He\,II}$.
From the incidence of Ly$\alpha$ and Ly$\beta$ transmission spikes we infer 
similar photoionization rates of 
$\Gamma _{\mathrm{He\,II}} = (2.0^{+0.6}_{-0.5})\times 10^{-15}\mathrm{\,s^{-1}}$ 
at $3.51 < z < 3.66$ and 
$\Gamma _{\mathrm{He\,II}} = (0.9\pm0.3)\times 10^{-15}\mathrm{\,s^{-1}}$ 
at $3.460 < z < 3.685$, respectively.
Because the transmission spikes indicate fully ionized regions at $z\gtrsim 3.5$ 
along both lines of sight, our observations provide further evidence that 
\ion{He}{2} reionization had substantially progressed at these redshifts.
\end{abstract}


\section{Introduction} \label{sec:intro}

The epoch of helium reionization, when helium lost its second electron, was the 
last major phase transition in the intergalactic medium (IGM). While hydrogen 
reionization was substantially complete by $z \sim 6$ 
\citep{Fan2006, Bosman2018, Eilers2018, PlanckCollab2018, Banados2018, 
Davies2018} and might have lasted until $z \sim 5.2$ 
\citep{Becker2015, Kulkarni2019b, NasirDAloisio2020, Keating2020b, Choudhury2021}, 
helium was fully ionized later at $z \sim 3$ due to the required hard UV photons 
($E \ge 54.4\mathrm{\,eV}$) that could only be provided by quasars 
\citep[e.g.,][]{MadauMeiksin1994, Fardal1998, Miralda-Escude2000, Sokasian2002, FurlanettoOh2008, McQuinn2009b, Compostella2013, Compostella2014}. 
The timing and spatial morphology of helium reionization is of great interest 
to observational cosmology, because it determines the spatial variations in 
amplitude and spectral shape of the high-redshift UV background 
\citep[e.g.,][]{FurlanettoDixon2010,Davies2017,Meiksin2020}, and it constrains 
the contribution of quasars to it \citep[e.g.,][]{Compostella2014, Garaldi2019a, Puchwein2019, Kulkarni2019}. 

In particular, extended photoheating during the helium reionization epoch 
governs the thermal evolution of the IGM at $z < 6$ 
\citep[e.g.,][]{McQuinn2009b, Compostella2013, Compostella2014, Onorbe2017, LaPlante2017, Puchwein2019}. 
Measurements of the IGM temperature at mean density obtained from the 
\ion{H}{1} Ly$\alpha$ forest with a variety of techniques show gradual heating 
at $z<4.5$ \citep{Becker2011,Boera2014,Boera2019} with a broad temperature 
maximum at $2.8\lesssim z \lesssim 3.4$ \citep[e.g.,][]{Ricotti2000,Schaye2000,Lidz2010,Boera2014,Hiss2018,Walther2019}. 
This temperature evolution is consistent with an extended \ion{He}{2} 
reionization epoch ending at $z\sim 3$ \citep{Puchwein2019}.

Direct evidence for delayed extended \ion{He}{2} reionization can be gained 
from intergalactic \ion{He}{2} Lyman series absorption against $z>2$ quasars 
in the FUV from space \citep[e.g.,][]{Miralda-Escude1993,MadauMeiksin1994,Jakobsen1994}.
However, the fraction of quasars showing emission at the \ion{He}{2} Ly$\alpha$ 
rest-frame wavelength $\lambda_\alpha=303.7822$\,\AA\ strongly decreases with increasing 
redshift due to the declining quasar luminosity function and increasing 
cumulative \ion{H}{1} Lyman continuum absorption \citep{MollerJakobsen1990,PicardJakobsen1993,WorseckProchaska2011}.
Wide-field FUV photometry obtained with the \textit{Galaxy Evolution Explorer} 
\citep[\textit{GALEX};][]{Martin2005,Morrissey2007} enabled the selection of 
likely \ion{He}{2}-transparent sightlines \citep{Syphers2009a,Syphers2009b,WorseckProchaska2011}. 
But only the FUV-brightest of these allow for efficient follow-up with 
\textit{HST}'s Cosmic Origins Spectrograph \citep[COS;][]{Green2012} at a data 
quality sufficient for quantitative measurements of intergalactic \ion{He}{2} 
absorption \citep[][hereafter \citetalias{Worseck2019}]{Worseck2011,Syphers2012,Worseck2016,Worseck2019}.

To date, science-grade (signal-to-noise ratio S/N$\gtrsim 3$) \textit{HST} 
spectra of 25 \ion{He}{2} sightlines sample the redshift evolution of the 
large-scale ($\sim 40$\ comoving Mpc) \ion{He}{2} Ly$\alpha$ absorption 
at $2.3\lesssim z\lesssim 3.8$ \citep[e.g.,][\citetalias{Worseck2019}]{Reimers1997,Heap2000,Shull2010,SyphersShull2013,SyphersShull2014,Worseck2011,Worseck2016}.
The small sightline-to-sightline variance in the measured effective optical 
depth $\tau_\mathrm{eff}$ at $z\lesssim 2.7$ is consistent with expectations 
from IGM density fluctuations and a spatially uniform UV background, marking 
the end of the \ion{He}{2} reionization epoch at $z\simeq 2.7$ (\citealt{Worseck2011}; 
\citetalias{Worseck2019}). At higher redshifts, the increasing 
$\tau _\mathrm{eff}$ variations are compatible with increasing UV background 
fluctuations in a still predominantly ionized IGM with a median volumetric 
\ion{He}{2} fraction of $x_\mathrm{He\,II}\simeq 0.025$ at $z\simeq 3.1$ 
(\citealt{Worseck2016}; \citealt{Davies2017}; \citetalias{Worseck2019}). 
However, beyond the tail end of \ion{He}{2} reionization, the constraints 
at $z>3.3$ are limited by the decreasing sample size and saturation in 
\ion{He}{2} Ly$\alpha$ at the sensitivity limit of \textit{HST}/COS \citepalias{Worseck2019}.

Higher-order Lyman series absorption can provide additional constraints on the 
reionization history due to higher saturation limits, 
e.g.\ for Ly$\beta$ $\tau_\beta(z)\simeq 0.16\tau_\alpha(z)$.
This has been exploited extensively to probe to higher \ion{H}{1} fractions 
at $z>5.5$ \citep{Becker2001, White2003, Fan2006, Gallerani2006, Becker2015, McGreer2015, Davies2018, Eilers2019, NasirDAloisio2020, Keating2020a, Yang2020}.
For the measured effective optical depths the above conversion does not apply, 
but instead depends on the IGM density structure 
\citep[e.g.][]{OhFurlanetto2005,Fan2006} 
and its thermal state \citep[e.g.][]{FurlanettoOh2009,Eilers2019}.
Moreover, Ly$\beta$ absorption is entangled with foreground Ly$\alpha$ 
absorption at $z_\mathrm{fg}=(1+z_\beta)\lambda_\beta/\lambda_\alpha-1$ that is 
best accounted for by appropriate forward-modeling \citep{Eilers2019,Keating2020a}.
Analogously, \ion{He}{2} Ly$\beta$ absorption enables one to improve the 
constraints on the \ion{He}{2} fraction by a factor of $\sim 3$ 
to $x_\mathrm{He\,II}\gtrsim 0.1$, particularly in underdense regions 
identified in the co-spatial \ion{H}{1} Ly$\alpha$ forest \citep{McQuinn2009}.
However, due to low UV instrument sensitivity in the wavelength range of 
interest, \ion{He}{2} Ly$\beta$ observations have been challenging 
\citep{Zheng2004, Syphers2011}. Additionally, the inferred \ion{He}{2} 
Ly$\alpha$ effective optical depths at $2.7\lesssim z\lesssim 3.4$ strongly 
depend on the uncertain modeling of foreground Ly$\alpha$ absorption, resulting 
in constraints similar to the direct \ion{He}{2} Ly$\alpha$ measurements \citep{Syphers2011}.

Besides the mean large-scale IGM absorption, several additional statistics have 
been used to probe the \ion{H}{1} reionization history with \ion{H}{1} 
absorption spectra, such as the length of continuous low-transmission 
regions \citep[so-called dark gaps,][]{SongailaCowie2002, Paschos2005, Fan2006, Gallerani2006, Gallerani2008, Gnedin2017} 
and the properties of isolated transmission spikes between these regions 
\citep{Gallerani2006, Gallerani2008, Gnedin2017, Barnett2017, Chardin2018, Garaldi2019, Gaikwad2020, Yang2020}.
While these statistics provide an intuitive feature decomposition of the 
high-redshift Ly$\alpha$ absorption, inferences on the reionization epoch 
require detailed forward-modeling of simulated spectra to match the spectral 
resolution and noise properties of the data (e.g., sky line residuals for dark gaps).
Extension to Ly$\beta$ is complicated by foreground Ly$\alpha$ modeling and 
numerical convergence in underdense regions, although the Ly$\beta$ dark gap 
distribution may discriminate between \ion{H}{1} reionization scenarios \citep{NasirDAloisio2020}.
Analogously, the \ion{He}{2} Ly$\alpha$ dark gap and transmission spike 
distributions have been suggested as powerful diagnostics to distinguish 
\ion{He}{2} reionization models \citep{Compostella2013}, but have not been 
rigorously applied due to limited high-quality data for just two well-studied 
\ion{He}{2} sightlines at $z\sim 2.8$ and $3.2$ \citep{Reimers1997,Heap2000,Smette2002,Shull2010,SyphersShull2014}.

Here we present new high-resolution ($R \sim 14,000$) \ion{He}{2} Ly$\alpha$ 
and Ly$\beta$ absorption spectra of the two FUV-brightest quasars 
at $z_\mathrm{em} > 3.5$, which have been discovered and analyzed at low 
spectral resolution by \citetalias{Worseck2019}. These two 
rare\footnote{\citet{WorseckProchaska2011} estimated that just 
$\sim 10$ $z_\mathrm{em}>3.5$ quasars are detectable with \textit{GALEX} 
at FUV$\lesssim 21.5$ on the full sky, and thus accessible to \textit{HST}.}
quasars, HE2QS\,J1630$+$0435 ($z _{\mathrm{em}} = 3.81$) and 
HE2QS\,J2311$-$1417 ($z _{\mathrm{em}} = 3.70$), are the only two quasars 
at $z_\mathrm{em} > 3.5$ that are bright enough to be observed with the 
high-resolution {\it HST}/COS G130M grating. The high-quality data and a new 
automated transmission spike measurement algorithm allow us to infer new 
constraints for the late stages of the \ion{He}{2} reionization epoch.
This paper is structured as follows. In Section~\ref{sec:obs_data_reduction} 
we describe the observations and our custom data reduction. 
In Section~\ref{sec:methods} we present our methods to detect \ion{He}{2} 
transmission spikes. Comparing the number of observed transmission spikes to 
outputs from numerical simulations we then infer the \ion{He}{2} photoionization 
rate (Section~\ref{sec:results}). Finally, we summarize in Section~\ref{sec:summary}.

We use a flat cold dark matter cosmology with dimensionless Hubble constant 
$h = 0.685$ ($H_{0} = 100h\mathrm{\,km\,s^{-1}\,Mpc^{-1}}$) and density parameters 
$(\Omega _{\mathrm{m}}, \Omega _{\mathrm{b}}, \Omega _{\mathrm{\Lambda}}) = (0.3, 0.047, 0.7)$, 
consistent with \citet{PlanckCollab2018}.

\section{Observations and Data Reduction} \label{sec:obs_data_reduction}

\subsection{{\it HST}/COS G130M Observations}

In {\it HST} Cycle 25 we observed HE2QS\,J1630$+$0435 and HE2QS\,J2311$-$1417 
with the COS G130M grating for $45,915\mathrm{\,s}$ and $54,920 \mathrm{\,s}$, 
respectively (Program 15356, PI Worseck). The observations were carried out 
at COS detector Lifetime Position 4 at three central wavelength settings 
($1222\mathrm{\,\AA}$, $1291\mathrm{\,\AA}$, $1327\mathrm{\,\AA}$) to provide 
a continuous wavelength coverage from $1070\mathrm{\,\AA}$ to 
$1470\mathrm{\,\AA}$ at a resolving power of $R = 10,000$--$17,000$ that varies 
with wavelength and the different settings. The observations were specifically 
scheduled to maximize the observation time spent in the Earth's shadow to 
reduce the contamination by geocoronal emission lines. Both targets were 
observed within $\sim 1$ month (except repeat observations due to guide star 
acquisition failures) to minimize the impact of quasar variability. 
Table~\ref{tab:obs_data} lists the successfully observed {\it HST} data sets. 

\begin{deluxetable}{llllc}
	\tablecaption{{\it HST}/COS G130M observational data.\label{tab:obs_data}}
	\tabletypesize{\scriptsize}
	\tablehead{
		\colhead{Object} & \colhead{Data set} & \colhead{Date} & \colhead{$t_{\mathrm{exp}} (\mathrm{s})$} & \colhead{$\lambda _{\mathrm{cent}} (\mathrm{\AA})$}}
	\startdata
	\object{HE2QS\,J1630$+$0435}	& \dataset[ldm601010]{http://archive.stsci.edu/cgi-bin/mastpreview?mission=hst&dataid=LDM601010} 	& 	2018 Jun 13 & 	5832 	&	1291 	\\
						& \dataset[ldm603010]{https://archive.stsci.edu/cgi-bin/mastpreview?mission=hst&dataid=LDM603010} 	& 	2018 Jul 8	&	11053	&	1327	\\
						& \dataset[ldm602010]{https://archive.stsci.edu/cgi-bin/mastpreview?mission=hst&dataid=LDM602010}	&	2018 Jul 11	&	11041	&	1222	\\
						& \dataset[ldm602020]{https://archive.stsci.edu/cgi-bin/mastpreview?mission=hst&dataid=LDM602020}	&	2018 Jul 11	&	2916	&	1291	\\
						& \dataset[ldm604010]{https://archive.stsci.edu/cgi-bin/mastpreview?mission=hst&dataid=LDM604010}	&	2018 Jul 13	&	2304	&	1327	\\
						& \dataset[ldm625010]{https://archive.stsci.edu/cgi-bin/mastpreview?mission=hst&dataid=LDM625010}	&	2018 Jul 24	&	5234	&	1291	\\
						& \dataset[ldm6h5020]{https://archive.stsci.edu/cgi-bin/mastpreview?mission=hst&dataid=LDM6h5020}	&	2018 Jul 31	&	2914	&	1327	\\
						& \dataset[ldm6h6010]{https://archive.stsci.edu/cgi-bin/mastpreview?mission=hst&dataid=LDM6h6010}	&	2018 Aug 13	&	2318	&	1291	\\
						& \dataset[ldm6h7010]{https://archive.stsci.edu/cgi-bin/mastpreview?mission=hst&dataid=LDM6h7010}	&	2018 Aug 21	&	2303	&	1327	\\	
	\object{HE2QS\,J2311$-$1417}	& \dataset[ldm605010]{http://archive.stsci.edu/cgi-bin/mastpreview?mission=hst&dataid=LDM605010} 	& 	2018 Sep 13 & 	12742 	&	1291 	\\
						& \dataset[ldm608010]{http://archive.stsci.edu/cgi-bin/mastpreview?mission=hst&dataid=LDM608010} 	& 	2018 Sep 14	&	10044	&	1327	\\
						& \dataset[ldm607010]{http://archive.stsci.edu/cgi-bin/mastpreview?mission=hst&dataid=LDM607010}	&	2018 Sep 15	&	10040	&	1222	\\
						& \dataset[ldm609010]{http://archive.stsci.edu/cgi-bin/mastpreview?mission=hst&dataid=LDM609010}	&	2018 Sep 22	&	7360	&	1327	\\
						& \dataset[ldm606010]{http://archive.stsci.edu/cgi-bin/mastpreview?mission=hst&dataid=LDM606010}	&	2018 Sep 27	&	12742	&	1291	\\
						& \dataset[ldm6h8010]{http://archive.stsci.edu/cgi-bin/mastpreview?mission=hst&dataid=LDM6h8010}	&	2018 Nov 5	&	1992	&	1327	\\
	\enddata
\end{deluxetable}

\subsection{Data Reduction} \label{sec:data_reduction}

The data were reduced using the \texttt{CALCOS} pipeline (v3.3.5) 
wrapped by our custom python code 
\texttt{FaintCOS}\footnote{\url{https://github.com/kimakan/FaintCOS}}. The code 
builds on the procedures of \citet{Worseck2016} and \citetalias{Worseck2019}, 
but with improved boxcar trace definition and flat-fielding. 
In contrast to \texttt{CALCOS}, \texttt{FaintCOS} accurately subtracts the 
dark current using dark frames, applies custom limits to the detector pulse 
height amplitude (PHA), adopts narrower extraction apertures that are adequate 
for point sources, and co-adds sub-exposures preserving the Poisson counts. 
\texttt{FaintCOS} provides a streamlined science-grade reduction of 
Poisson-limited \textit{HST}/COS FUV spectra in a single software environment. 
Reduction parameters can be tailored to the science objectives. 
\texttt{FaintCOS} is described in detail in Appendix \ref{sec:faintcos}.

In addition, we checked the science extraction apertures for occasional 
transient detector hotspots with a typical width of a few pixels that might 
mimic \ion{He}{2} transmission spikes. We identified these hotspots in stacked 
dark frames taken around the same time as the science data. The corresponding 
detector positions were masked in the science data, because hotspots cannot be 
removed with our dark current subtraction procedure. Although their overall 
effect is small for our data taken at several central wavelengths and 
focal-plane offset positions, custom masking of hotspots is required for the 
faintest COS targets such as ours.

Because of the expected narrow flux spikes in our data, wavelength alignment of 
different COS setups is essential. Comparing spectra of the calibration star 
AV\,75 (Program 15385) taken at different central wavelengths, we found a 
distortion in the \texttt{CALCOS} wavelength calibration of up to $0.2$\,\AA\ 
(2 resolution elements) at the short wavelength edges (300 pixels) of both FUV 
detector segments. This is likely due to an imperfect correction of geometrical 
distortion. We did not manually flag these regions because there is no 
significant flux detected, neither in individual \textit{HST} data sets, 
nor in the co-added spectra. Conversely, current COS data showing unsaturated 
absorption lines in these regions should be checked carefully.

Custom PHA ranges specific to the science data allow for the suppression of 
the detector dark current \citep{Worseck2016}. From the PHA distributions of 
geocoronal emission lines and the quasar continuum redward of \ion{He}{2} 
Ly$\alpha$ we inferred $2 \leqslant \mathrm{PHA} \leqslant 12$. Our custom 
boxcar extraction apertures and PHA limits result in a $\sim 60$\% reduction 
of the dark current with respect to the default \texttt{CALCOS} procedure. 

The dark current in the science extraction apertures was estimated from 
post-processed dark frames taken within two months around the observation date.
To match the orbital environmental conditions during the science observations, 
the dark frames were restricted to those with a Kolmogorov–Smirnov test 
statistic $D<0.03$ between the respective PHA distributions in unilluminated 
detector areas (Appendix~\ref{sec:faintcos}). The $D$ limit was increased until 
at least 5 matching dark frames were found. The exposures at the central 
wavelength 1222\,\AA\ had been taken at a higher voltage level on detector 
segment B, which had not been included in the COS dark monitoring 
programs 14940 and 15533. Because the dark current depends on the detector 
voltage, we acquired specific dark frames at the matching segment B voltage 
to ensure a percent-level accurate dark subtraction for these 
data\footnote{Reduction with the standard dark monitoring data resulted in 
a 10\% overestimation of the dark current, so dark frames matching the detector 
voltage of the science data are required for an accurate dark 
current subtraction.}. Our improved dark current subtraction method has a 
negligible systematic error of $\le 0.5$\% and a random error of 
$\lesssim 5$\% (Appendix~\ref{sec:dc_validation_test}).

Although the dark current is the dominant background component, the diffuse UV 
sky background and scattered geocoronal Ly$\alpha$ emission may not be 
negligible \citep{Worseck2016}. The diffuse UV sky background was subtracted by 
using \textit{GALEX} data \citep{Murthy2014} as described 
in \citet{Worseck2016}. The subtracted fluxes are small 
($f_{\lambda,\mathrm{sky}}\sim 10^{-18}\mathrm{\,erg\,cm^{-2}\,s^{-1}\,\AA ^{-1}}$), 
and correspond to 2--12\% of the estimated dark current. Additionally, we 
inspected both co-added spectra for resolved H$_2$ fluorescence 
lines \citep{Sternberg1989} which can mimic \ion{He}{2} transmission spikes. 
Since even the strongest lines are not present, we conclude that H$_2$ 
fluorescence is negligible along these two sightlines and that the sky 
background is continuous.

Contamination by geocoronal emission lines was suppressed by considering data 
taken during orbital night (Sun altitude $\le 0\degr$) in the affected 
wavelength ranges. During orbital night, \ion{O}{1}~$\lambda 1304$ and 
\ion{N}{1}~$\lambda 1200$ are very weak or even negligible, depending on the 
geomagnetic latitude, solar activity, and \textit{HST}'s orientation to the Sun.
The affected spectral regions in the co-added data were replaced with night-only 
data to suppress these lines below the detection limit 
(Appendix~\ref{sec:geocor_decont}). Additionally, we checked for scattered 
geocoronal Ly$\alpha$ emission by comparing the day-only and night-only fluxes 
in four saturated regions close to Ly$\alpha$. The fluxes are either 
statistically insignificant ($f_\lambda < 3\times 10^{-18}\,\mathrm{erg\,cm^{-2}\,s^{-1}\,\AA ^{-1}}$ 
at $2\sigma$) or within the systematic error of the dark current,
such that scattered light is considered negligible.

The sub-exposures per object were co-added and calibrated in count space to 
preserve their Poisson statistics, binning to $0.04$\,\AA\,pixel$^{-1}$, 
corresponding to 2--3 pixels per resolution element. Because the resolution 
varies with wavelength and G130M central wavelength, we quote for the merged 
spectra $R\sim 14,000$ at $\lambda\sim 1300$\,\AA. Statistical Poisson errors 
for the background-subtracted flux were calculated using the approach of \citet{FeldmanCousins1998}. 
The resulting Poisson S/N is $\sim 3$ per pixel in the continuum near 
\ion{He}{2} Ly$\alpha$ in the quasar rest frame, but strongly varies with 
wavelength due to the different pixel exposure times and \ion{He}{2} absorption.

\subsection{Continuum Definition}

To correct the co-added spectra for Galactic extinction we used the 
line-of-sight selective extinction $E(B - V)$ from \citet{Schlegel1998} and 
the extinction curve derived by \citet{Cardelli1989} assuming the Galactic 
average ratio between the total $V$ band extinction and selective 
extinction $R_V=3.1$.

Since both G130M spectra cover only a short part of the quasar continuum 
redward of \ion{He}{2} Ly$\alpha$, which makes continuum fitting impossible, 
we used the power-law continua derived by \citetalias{Worseck2019}. 
In order to correct for intrinsic quasar variability between the two 
observational epochs, we re-observed each quasar with the G140L grating during 
one of the G130M visits. Exposure times were $2,182$\,s for HE2QS\,J1630$+$0435 
and $2,558$\,s for HE2QS\,J2311$-$1417, respectively. We reduced the G140L data 
with the same techniques as we did for the G130M exposures and additionally 
corrected for scattered geocoronal Ly$\alpha$ emission \citep{Worseck2016}.
In the G140L spectra we calculated the mean flux redward of 
\ion{He}{2} Ly$\alpha$ emission in 20\,\AA\ bins (1430--1730\,\AA\ for 
HE2QS\,J2311$-$1417 and 1460--1730\,\AA\ for HE2QS\,J1630$+$0435). 
For both quasars the flux ratio between the epochs does not depend on wavelength.
We adopted the mean flux ratios and their standard errors to scale the 
\citetalias{Worseck2019} continua and their $1\sigma$ uncertainties 
(Table~\ref{tab:flux_factors}).

\begin{deluxetable}{lllc}
	\tablecaption{Flux Ratio for G140L Spectra at Different Epochs.\label{tab:flux_factors}}
	\tablehead{
		\colhead{Object} & \colhead{1st Epoch} & \colhead{2nd Epoch} & \colhead{Mean Flux Ratio} \\[-2ex]
		\colhead {}    & \colhead {}           & \colhead {} & \colhead {(2nd/1st epoch)}}
	\startdata
	HE2QS\,J2311$-$1417 	& 	2015 Nov 7 & 	2018 Sep 22	&	$1.02\pm 0.03$ 	\\
	HE2QS\,J1630$+$0435 	& 	2013 Apr 12	&	2018 Jul 31	&	$0.87\pm 0.02$	\\
	\enddata
\end{deluxetable}

\section{Two High-resolution \ion{He}{2} Absorption Spectra Probing $z > 3.5$}
\label{sec:data_reduction_results}

\begin{figure*}
	\includegraphics[width = \textwidth]{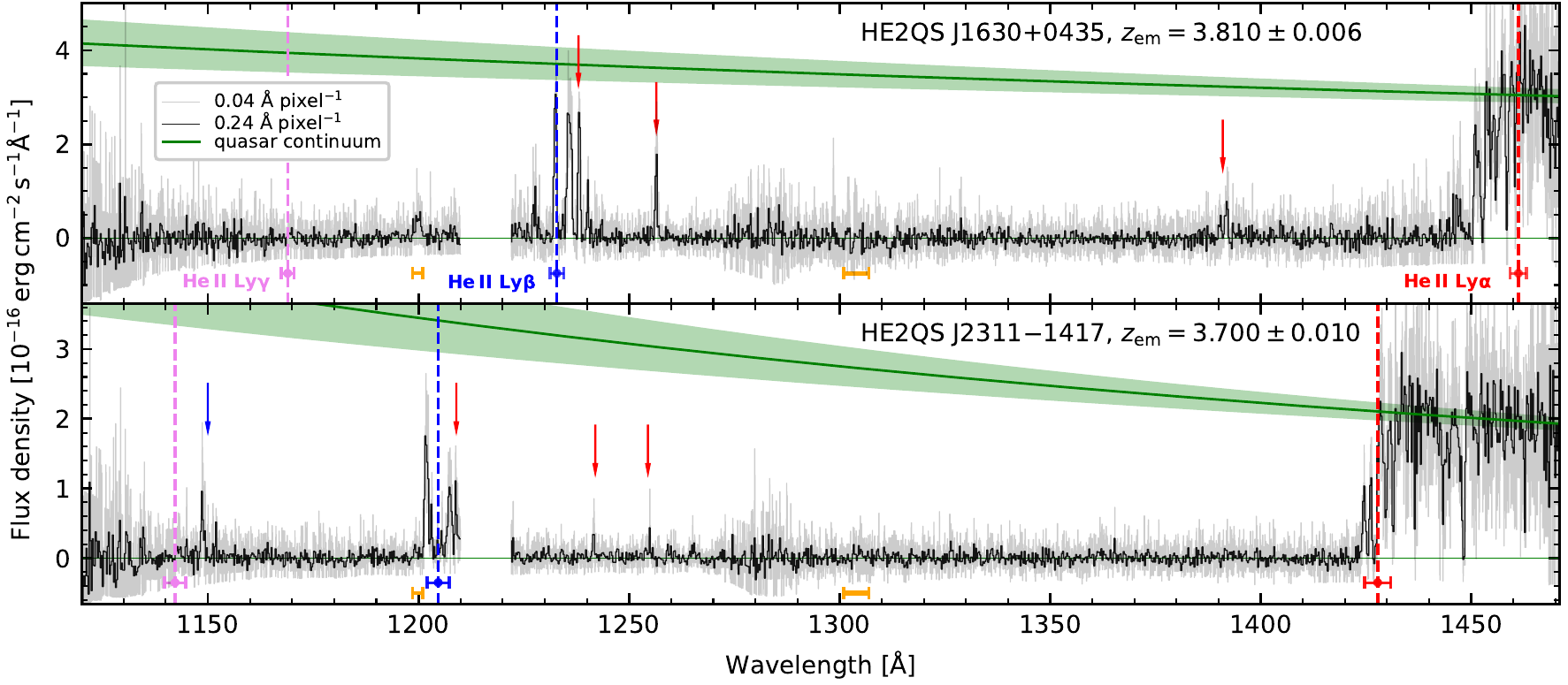}
	\caption{Extinction-corrected {\it HST}/COS G130M spectra of our 
	two \ion{He}{2}-transparent quasars sampled at $0.04\,$\AA\,pixel$^{-1}$ 
	(2--3 pixels per resolution element, gray) and $0.24\,$ \AA\,pixel$^{-1}$\, 
	(undersampled, black). The green line shows the best-fit power law 
	continuum with $1\sigma$ error determined by \citetalias{Worseck2019} and 
	corrected for quasar variability between the observation epochs. 
	The vertical dashed lines show \ion{He}{2} Ly$\alpha$ (red), 
	Ly$\beta$ (blue) and Ly$\gamma$ (violet) in the quasar rest frame with 
	indicated $1\sigma$ quasar redshift error. Visible \ion{He}{2} Ly$\alpha$ 
	and Ly$\beta$ transmission spikes are marked with red and blue arrows, 
	respectively. The spectral region contaminated by geocoronal Ly$\alpha$ 
	emission at 1216\,\AA\ is not shown. The orange horizontal lines indicate 
	the spectral ranges around geocoronal \ion{N}{1}\,$\lambda$1200 
	and \ion{O}{1}\,$\lambda$1304 where we use only data taken in orbital night.}
	\label{fig:reduced_spectra}
\end{figure*}

\subsection{General Overview} \label{sec:general_description}

Figure~\ref{fig:reduced_spectra} shows the reduced G130M spectra of both 
quasars with $0.04$\,\AA\,pixel$^{-1}$ and $0.24$\,\AA\,pixel$^{-1}$ binning 
(under-sampled for visualization). Both spectra cover a short wavelength range 
of the quasar continuum longward of \ion{He}{2} Ly$\alpha$ in the quasar 
rest frame. The quasar redshifts have been measured by \citet{Khrykin2019}, 
using the \ion{C}{4} and H$\beta$ emission lines for HE2QS\,J2311$-$1417 and 
HE2QS\,J1630$+$0435, respectively. The residual flux immediately shortward of 
\ion{He}{2} Ly$\alpha$ and \ion{He}{2} Ly$\beta$ ($\lambda_\beta=256.317$\,\AA) 
is due to the highly ionized \ion{He}{2} quasar proximity zones \citep{Khrykin2019}, 
which were excluded from further analysis.

The remaining spectral range is covered by predominantly saturated 
intergalactic \ion{He}{2} Lyman series absorption. Both sightlines show 
long \ion{He}{2} Ly$\alpha$ Gunn-Peterson troughs \citep{GunnPeterson1965} 
with occasional transmission features (i.e., at 1255\,\AA\ and 1390\,\AA\ 
toward HE2QS\,J1630$+$0435). The enhanced scatter at 1270\,\AA$<\lambda<$1290\,\AA\ 
is due to the short exposure time in the G130M 1222\,\AA\ setup covering 
this wavelength range. The spectral range shortward of \ion{He}{2} Ly$\beta$ in 
the quasar rest frame shows overlapping intergalactic high-redshift 
\ion{He}{2} Ly$\beta$ and low-redshift foreground \ion{He}{2} Ly$\alpha$ 
absorption. Due to the increasing contamination and low S/N we excluded 
the \ion{He}{2} Ly$\gamma$ troughs from further analysis.

\subsection{Resolved \ion{He}{2} Transmission Spikes}

The strongest \ion{He}{2} Ly$\alpha$ transmission spikes in both spectra are 
already known from the low-resolution data \citepalias{Worseck2019}, 
but are now resolved (FWHM $\gtrsim 50\mathrm{\,km\,s^{-1}}$). Furthermore, our 
deep G130M data enable us to verify very weak spikes previously only 
tentatively detected in the G140L spectra, i.e., at $\sim1250$\,\AA\ in 
the HE2QS\,J2311$-$1417 spectrum (see Section~\ref{sec:transmission_in_lya}). 
For the first time, we detect intergalactic $z > 3.4$  \ion{He}{2} Ly$\beta$ 
spikes at $\sim1150\mathrm{\,\AA}$ (HE2QS\,J2311$-$1417) and possibly 
at $\sim1200\mathrm{\,\AA}$ (HE2QS\,J1630$+$0435), which are further analyzed 
in Section~\ref{sec:tranmission_in_lyb}.

\begin{figure}[t]
	\includegraphics[width=\columnwidth]{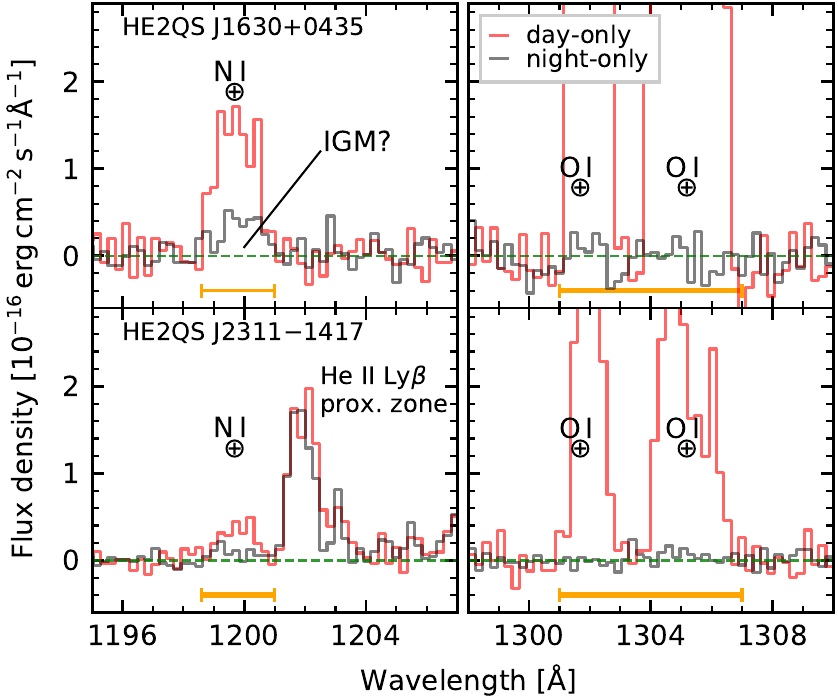}
	\caption{Comparison between day-only (red) and night-only (gray) data in 
	spectral regions contaminated by geocoronal \ion{N}{1} and \ion{O}{1} for 
	HE2QS\,J1630$+$0435 (upper panels) and HE2QS\,J2311$-$1417 (lower panels). 
	The horizontal lines indicate spectral ranges replaced by night-only data.}
	\label{fig:day_night_comparison}
\end{figure}

The flux spikes at $\sim 1200\mathrm{\,\AA}$ in the spectrum of 
HE2QS\,J1630$+$0435 lie suspiciously on top of the geocoronal \ion{N}{1} 
$\lambda1200$ line. Therefore, we closely investigated the accuracy of the 
geocoronal decontamination. Figure~\ref{fig:day_night_comparison} shows the 
night-only and day-only spectra for both quasars at geocoronal \ion{O}{1} 
$\lambda$1304 and \ion{N}{1} $\lambda$1200. The \ion{O}{1} $\lambda1304$ lines 
show no residual flux above the detection limit 
at $1301\mathrm{\,\AA}$--$1307\mathrm{\,\AA}$ in both night-only spectra.
However, the weaker \ion{N}{1} $\lambda1200$ line still seems to be present in 
the night-only spectrum of HE2QS\,J1630$+$0435 which is 
very unusual \citepalias{Worseck2019}. In contrast, geocoronal \ion{N}{1} in 
the spectrum of HE2QS\,J2311$-$1417 vanishes as expected, while the strong 
spike in the \ion{He}{2} Ly$\beta$ proximity zone next to it does not change.
Analysis of blank-sky {\it HST}/COS observations 
(Appendix~\ref{sec:geocor_decont}) could not definitely determine the origin of 
the flux at 1200\,\AA\ in the HE2QS\,J1630$+$0435 spectrum. Therefore, we 
decided to exclude it from further analysis.

\subsection{Comparison with the \textit{HST}/COS G140L Spectra}

We compared the G130M spectra to the low-resolution G140L data analyzed 
by \citetalias{Worseck2019}. The most significant differences in the data 
reduction are that G140L spectra must be corrected for scattered geocoronal 
Ly$\alpha$ emission unlike our G130M data (Section~\ref{sec:obs_data_reduction}), 
and that \citetalias{Worseck2019} neglected flat-fielding, which is 
inconsequential in the wavelength range of interest.

\begin{figure*}[h]
	\includegraphics[width=\textwidth]{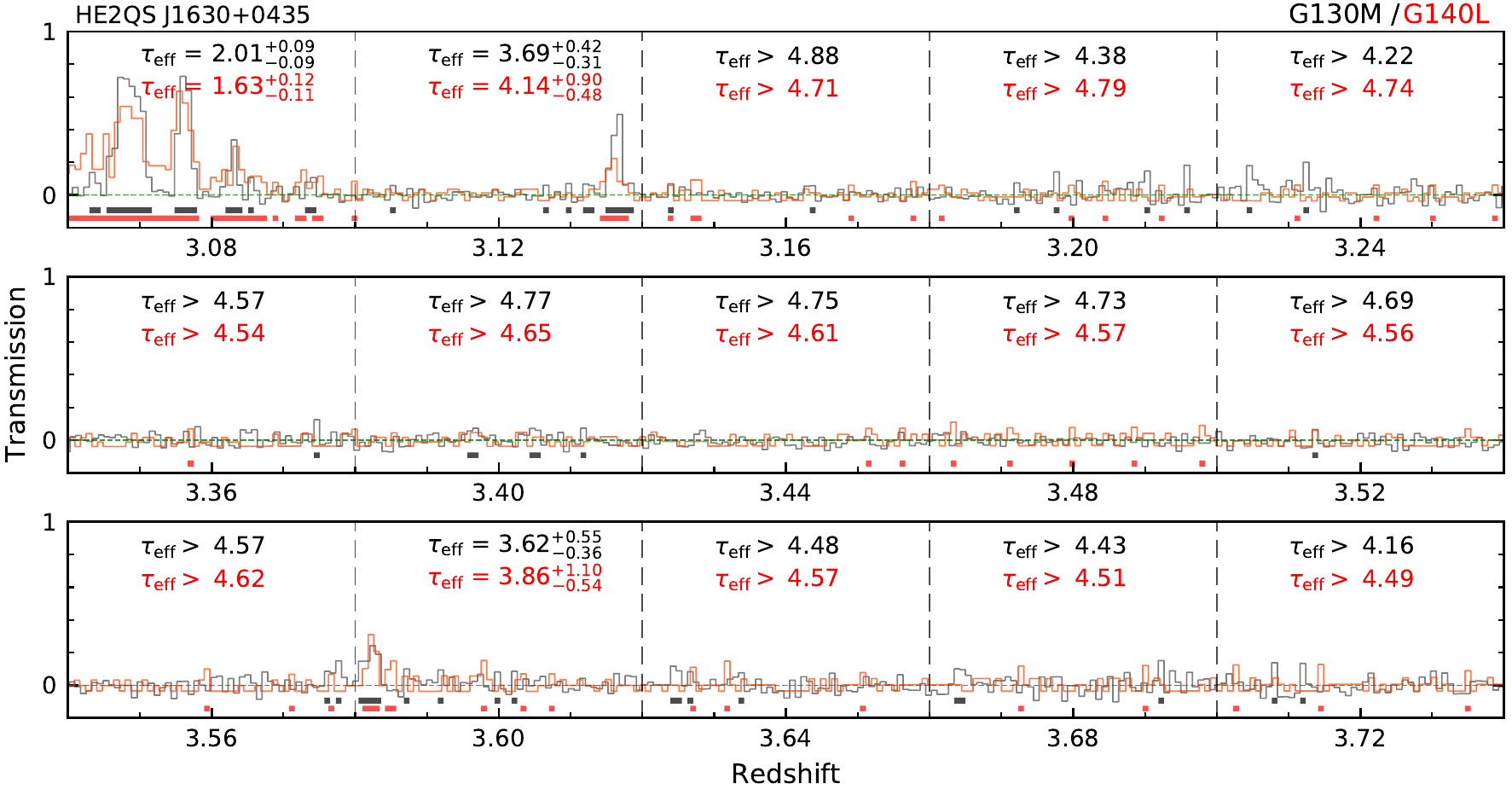}
	\caption{High-resolution G130M (gray) and low-resolution G140L 
	(red, \citetalias{Worseck2019}) Ly$\alpha$ transmission spectra of 
	HE2QS\,J1630$+$0435 binned to $0.24\,$\AA\,pixel$^{-1}$. The thick red and 
	black points and lines underneath the zero line (dotted line) indicate 
	pixels with Poisson probability $P(>N|B)<0.0227$, i.e.\ pixels with 
	transmission detected at $>2\sigma$ significance. We also indicate the 
	effective optical depths measured in $\Delta z = 0.04$ intervals 
	(vertical lines) with $2\sigma$ double-sided Poisson error bars, or as 
	$2\sigma$ lower limits if $P(>N|B)\ge 0.0227$.
	}
	\label{fig:opt_depth_j1630}
\end{figure*}
\begin{figure*}[h]
	\includegraphics[width=\textwidth]{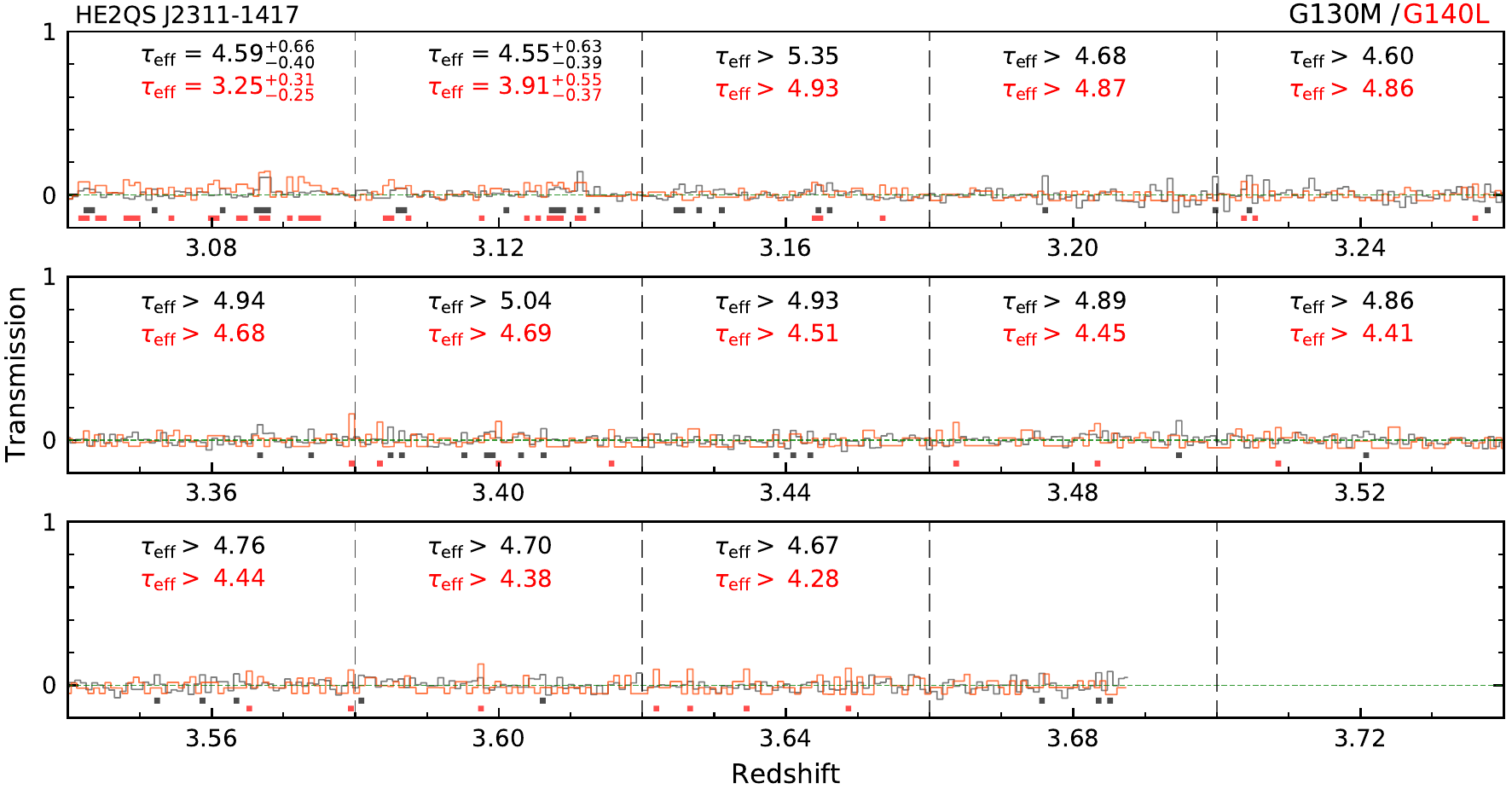}
	\caption{Similar to Figure~\ref{fig:opt_depth_j1630} but for HE2QS\,J2311$-$1417.}
	\label{fig:opt_depth_j2311}
\end{figure*}

Figures~\ref{fig:opt_depth_j1630} and \ref{fig:opt_depth_j2311} show the 
normalized G130M and G140L \ion{He}{2} Ly$\alpha$ transmission spectra using 
the same binning of $0.24$\,\AA\,pixel$^{-1}$. The strongest spikes previously 
detected in the G140L spectra are confirmed in the G130M spectra, e.g.\ at 
$z\sim 3.08$ and $z \sim 3.135$ in Figure~\ref{fig:opt_depth_j1630}. Due to the 
higher effective spectral resolution of the undersampled G130M data, the spikes 
are more prominent than in the G140L data, e.g.\ the spike at $z \simeq 3.135$ 
in Figure~\ref{fig:opt_depth_j1630}. In order to quantify the significance of 
the measured \ion{He}{2} Ly$\alpha$ transmission, we calculated for each pixel 
the probability 
\begin{equation}\label{eq:prob}
P(> N|B) = 1 - \sum_{k=0}^{N}\frac{B^{k} e^{-B}}{k!}
\end{equation}
that the detected $N$ Poisson counts are consistent with the estimated 
background $B$. Very low $P$ values indicate statistically significant 
\ion{He}{2} transmission ($N \gg B$). In Figures~\ref{fig:opt_depth_j1630} and 
\ref{fig:opt_depth_j2311} we have marked pixels with $P(>N|B) < 0.0227$ that 
correspond to a $>2\sigma$ detection.

From the total 860 $0.24$\,\AA\ pixels at $3.06 \le z \le 3.74$ in the spectrum 
of HE2QS\,J1630$+$0435 we would expect $\sim 20$ random pixels with 
$P(>N|B) < 0.0227$ from the Poisson noise alone. Therefore, we regard individual 
pixels as not reliable to find real transmission. Nevertheless, two or more 
consecutive pixels with $P(>N|B) < 0.0227$ are a good indicator for real 
transmission spikes, because the probability to find two such pixels arising 
just from Poisson noise is only $\sim 0.07$\%. Due to the varying exposure time 
and Poisson noise, some regions show apparent transmission, e.g.\ at 
$z = 3.20$--$3.24$ in the G130M data (Section~\ref{sec:general_description}).
The G130M data show that some of the apparent unresolved spikes in the G140L 
data are caused by background Poisson noise, e.g.\ $z\sim 3.60$ in 
Figure~\ref{fig:opt_depth_j2311}). Another advantage of the G130M data, besides 
the higher resolution, is the fact that the contamination from scattered 
geocoronal Ly$\alpha$ emission is negligible. This is especially noticeable in 
the regions close to geocoronal Ly$\alpha$ where the G140L spectra show 
systematically higher transmission due to residual scattered light, i.e.\ at 
$z<3.10$ in Figure~\ref{fig:opt_depth_j2311}.

\begin{figure}
	\includegraphics[width=\columnwidth]{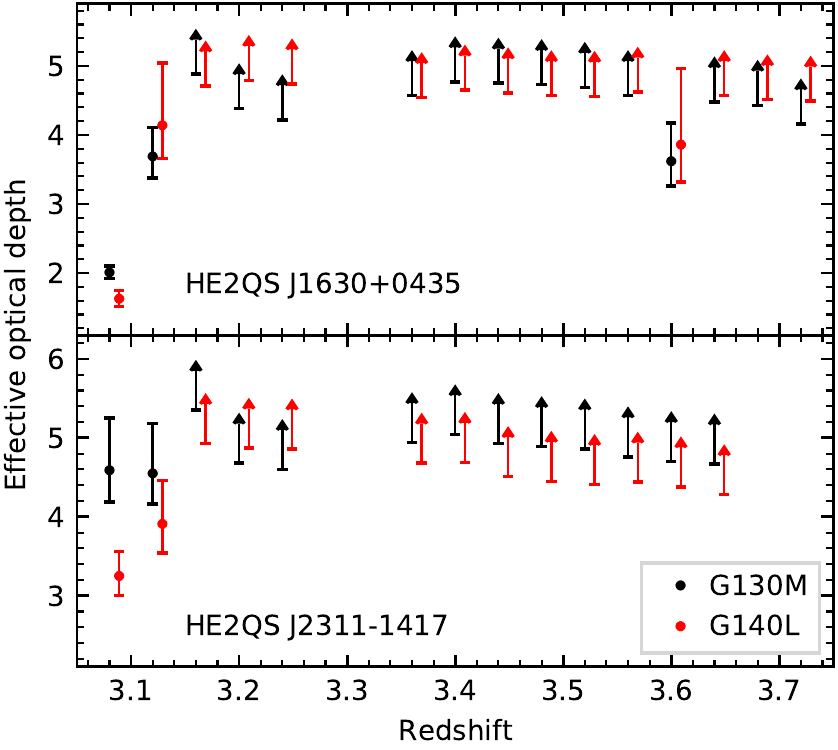}
	\caption{Comparison of the \ion{He}{2} Ly$\alpha$ effective optical depths 
	measured in G140L (red) and G130M (black) spectra with double-sided 
	$2\sigma$ Poisson errors or $2\sigma$ lower limits (arrows). The redshift 
	range $3.26<z<3.34$ is not reliable due to possible residual geocoronal 
	\ion{O}{1} contamination in the G140L data.}
	\label{fig:tau_eff_j1630}
\end{figure}

Additionally, we measured the Ly$\alpha$ effective optical depth 
$\tau_\mathrm{eff}=-\ln\langle f_\lambda/E_\lambda\rangle$, where $f_\lambda$ 
is the quasar flux density corrected for Galactic extinction, $E_\lambda$ is 
the respective extrapolated continuum, and $\langle\rangle$ denotes the average 
taken over a redshift range $\Delta z$. For comparison we adopted the same 
technique and the same redshift intervals $\Delta z = 0.04$ as 
\citetalias{Worseck2019}. In short, we maximized the Poisson likelihood function
\begin{equation}
\label{eq:poisson-like}
L = \prod_{j=1}^{n}\frac{(S_{j} + B_{j})^{N_{j}} e^{-(S_{j} + B_{j})}}{N_{j}!}
\end{equation}
for $n$ contiguous pixels in the $\Delta z = 0.04$ bin with $N_{j}$ registered 
counts, the background $B_{j}$ and the not yet determined non-integer source 
counts $S_{j} = t_{j}C_{j}K_{\mathrm{FF},j}E_{j}e^{-\tau _{\mathrm{eff}}}$ with 
the pixel exposure time $t_{j}$, the extinction-corrected  flux calibration 
curve $C_{j}$, the flat-field correction $K_{\mathrm{FF},j}$ and the 
continuum $E_{j}$. Figure~\ref{fig:tau_eff_j1630} shows the measured 
$\tau_\mathrm{eff}$ in the G130M and G140L spectra with $2\sigma$ error bars 
and $2\sigma$ lower limits\footnote{\citetalias{Worseck2019} used $1\sigma$ 
errors and lower limits.} in case $P\ge 0.0277$ in the $\Delta z=0.04$ bin.
Except for the bin $3.06 < z < 3.10$ where the G140L data are affected by 
residual scattered geocoronal Ly$\alpha$ emission, the measurements in 
high-resolution and low-resolution data are in very close agreement. The G130M 
data have a comparable or higher sensitivity to high $\tau_\mathrm{eff}$ values 
than the G140L data.

In summary, the G130M data are overall more reliable than the G140L data 
according to the $\tau_\mathrm{eff}$ measurements. Furthermore, their higher 
resolution enables us to detect narrow transmission spikes at high significance 
to distinguish them from Poisson noise.

\section{\ion{He}{2} Transmission Spikes in Realistic Mock Spectra}
\label{sec:methods}

\subsection{\ion{He}{2} Mock Spectra from a Hydrodynamical Simulation}

\subsubsection{Simulation and UV Background Models}

To constrain the \ion{He}{2} reionization history with \ion{He}{2} transmission 
spikes we must compare our data to realistic mock spectra from a numerical 
simulation with a given \ion{He}{2} photoionization rate 
$\Gamma_\mathrm{He\,II}$, or equivalently with a given \ion{He}{2} fraction 
$x_\mathrm{He\,II}$. 
We used $\Delta z=0.04$ long skewers from \citetalias{Worseck2019}, which had 
been created from outputs of a cubic $(146\mathrm{\,cMpc})^{3}$ hydrodynamical 
simulation performed with the \texttt{Nyx} code \citep{Almgren2013, Lukic2015} 
applying photoionization and photoheating rates from \citet{HaardtMadau2012}.
The skewers had been made for the \citetalias{Worseck2019} effective optical 
depth measurements on a length scale $\Delta z=0.04$ (34\,cMpc at $z = 3.5$) 
in the redshift range $2.56 \le z \le 3.88$, and were initially 
longer ($\Delta z = 0.08$) to account for the low-resolution G140L line spread 
function \citepalias{Worseck2019}. At each redshift we used only the 
central $\Delta z=0.04$ of the available 1000 skewers with a pixel size 
of $2$--$3\mathrm{\,km\,s^{-1}}$.
The \ion{He}{2} Ly$\alpha$ optical depths were rescaled according to the 
UV background models as $\tau_\alpha \propto \Gamma _{\mathrm{He\,II}}^{-1}$ 
in the optically thin limit, which approximately holds at the tail end of 
the \ion{He}{2} reionization epoch. 

We infer photoionization rates from our observed data by using 
the predictions from a set of spatially uniform 
UV background models with different amplitudes. We created 1000 synthetic 
spectra for \ion{He}{2} Ly$\alpha$ and \ion{He}{2} Ly$\beta$ separately for a 
set of constant photoionization rates
($ 10^{-16.4}\,\mathrm{s^{-1}} \le \Gamma _{\mathrm{He\,II}} \le 10^{-13.3}\,\mathrm{s^{-1}}$ 
with step size $\Delta [\log (\Gamma_{\mathrm{He\,II}} /\mathrm{s^{-1}})] = 0.1$) 
by concatenating the $\Delta z = 0.04$ skewers.
Additionally, we used a spatially fluctuating UV background model 
\citep{Davies2017} for comparison, and to model foreground 
$z_\mathrm{fg}<3$ \ion{He}{2} Ly$\alpha$ 
absorption that overlaps with the high-redshift Ly$\beta$ 
absorption (Fig.~\ref{fig:reduced_spectra}).
This model was calculated in a $(500\mathrm{\,cMpc})^{3}$ volume with grid 
cells of $(7.8\,\mathrm{cMpc})^3$ using an analytic IGM absorber model and 
a quasar luminosity function, resulting in a spatially varying mean free 
path that increases from $\sim 20$\,cMpc at $z=3.2$ to $\sim 40$\,cMpc at $z=2.8$.
The usage of this specific model is justified by its excellent 
reproduction of the observed large-scale variations in the $z<3.3$ 
\ion{He}{2} Ly$\alpha$ effective optical depths \citepalias{Worseck2019}. 
Applying it to the outputs of the $(146\mathrm{\,cMpc})^{3}$ hydrodynamical 
simulation, we sufficiently capture \ion{He}{2} transmission features sourced 
by the density field and by UV background fluctuations. By doing so, we lose the 
correlation between the density field and radiation field, however, due to the 
rarity of quasars and long mean free path the correlation is expected to be 
fairly weak.

\subsubsection{Mock Spectra for \ion{He}{2} Ly$\alpha$}
\label{sec:mocks_lya}

\begin{figure}[t]
	\includegraphics[width=\columnwidth]{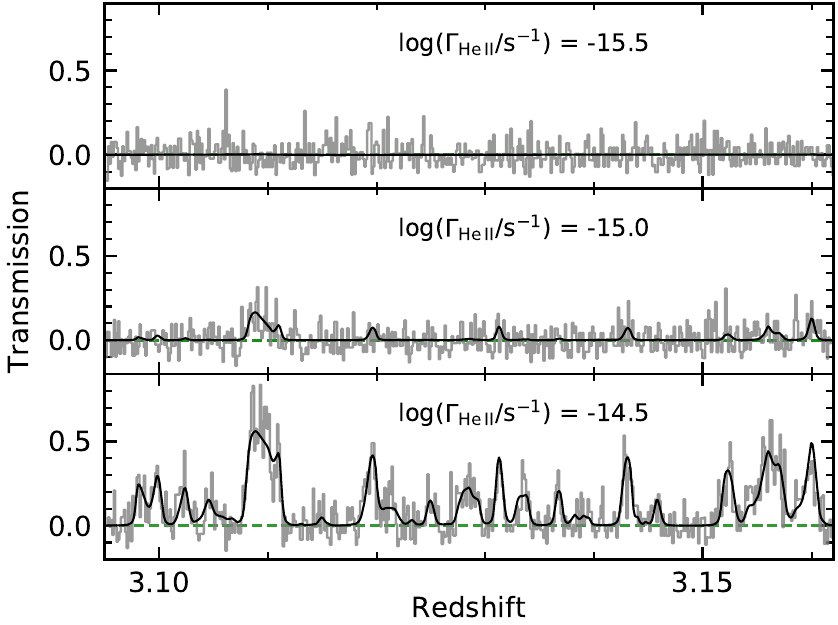}
	\caption{The effect of the photoionization rate on the number of spikes 
	in \ion{He}{2} Ly$\alpha$ mock spectra. The panels show the simulated 
	\ion{He}{2} transmission at G130M resolution with (gray) and without (black) 
	Poisson noise for the same skewer.}
	\label{fig:growing_forest}
\end{figure}

First, for any given \ion{He}{2} photoionization rate, random synthetic spectra 
were concatenated to 1000 long \ion{He}{2} Ly$\alpha$ transmission spectra 
covering the common redshift range $3.06\le z_\alpha\le 3.66$. The redshift 
ranges covered by only one of the sightlines are too short for the analysis.
The concatenation does not affect the final statistical comparison of the 
transmission spikes in smaller redshift bins $\Delta z = 0.15$ that account for 
possible redshift evolution of the \ion{He}{2} photoionization rate. Density 
discontinuities at the edges of the concatenated skewers are not of a concern 
due to the strong absorption. We excluded skewers with \ion{H}{1} Ly$\alpha$ 
optical depths $\tau _\mathrm{H\,I} > 3000$ that correspond to \ion{H}{1} 
column densities $N_{\mathrm{H\,I}} \gtrsim 10^{17.1}\mathrm{\,cm^{-2}}$, 
because a lack of strong \ion{H}{1} Lyman limit systems is required to 
render \ion{He}{2} observable.

Realistic COS mock spectra matching the characteristics of the observed spectra 
were produced as in \citetalias{Worseck2019}. First, we applied the 
{\it HST}/COS G130M line-spread function to the synthetic \ion{He}{2} Ly$\alpha$ 
transmission spectra. We used the exposure time weighted average {\it HST}/COS 
line-spread function at 1300\,\AA\ corresponding to $R \sim 14,000$, which is 
sufficient at our low Poisson S/N$\lesssim 3$. Then, the synthetic spectra were 
rebinned to the wavelength grid of the observed spectra. Next, we used the 
calibration curve, the pixel exposure time and the background model of the 
observed spectra to convert the \ion{He}{2} transmission to expected counts. 
Finally, the COS counts were simulated by adding Poisson noise according to the 
expected counts per pixel. To incorporate the estimated systematic background 
subtraction error, the background model of the observed spectra was varied 
assuming a Gaussian distribution with a standard deviation according to the 
estimated systematic error.

Figure~\ref{fig:growing_forest} shows the effect of the photoionization rate on 
the \ion{He}{2} Ly$\alpha$ transmission in mock spectra of the same density 
skewer. Generally, \ion{He}{2} transmission features arise in IGM 
underdensities \citep{Croft1997}. A higher photoionization rate results in 
stronger and more numerous spiky features from the emerging 
\ion{He}{2} Ly$\alpha$ forest. It also shows that the smallest features 
disappear in the noise as in the observed spectra.

\subsubsection{Mock Spectra for \ion{He}{2} Ly$\beta$}

\begin{figure}
	\includegraphics[width=\columnwidth]{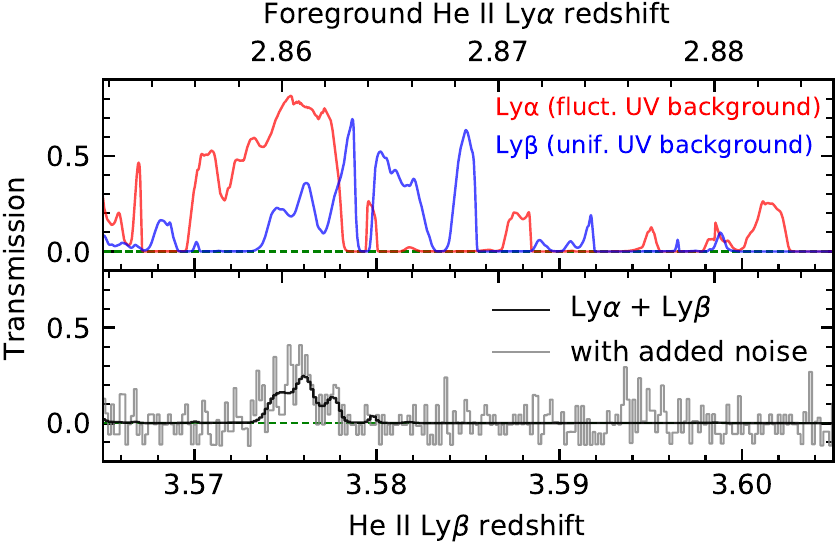}
	\caption{Creation of \ion{He}{2} Ly$\beta$ mock spectra. 
	\emph{Upper panel:} Two random $\Delta z = 0.08$ \ion{He}{2} transmission 
	skewers for high-redshift Ly$\beta$ ($z_\beta\sim 3.58$, blue) and 
	low-redshift Ly$\alpha$ ($z_\mathrm{fg}\sim 2.87$, red). 
	\emph{Lower panel:} Overlapped, i.e.\ multiplied \ion{He}{2} Ly$\alpha$ and 
	Ly$\beta$ transmission with (gray) and without (black) Poisson noise.}
	\label{fig:mock_creation}
\end{figure}

The \ion{He}{2} Ly$\beta$ forest overlaps with foreground \ion{He}{2} Ly$\alpha$ 
absorption at $z_\mathrm{fg}=(1+z_\beta)\lambda_\beta/\lambda_\alpha-1$.
Pure \ion{He}{2} Ly$\beta$ skewers were created by rescaling the optical depths 
of the \ion{He}{2} Ly$\alpha$ skewers by the respective rest frame wavelengths 
and oscillator strengths as
\begin{equation}
    \tau_\beta(z_\beta) = \frac{\lambda_\beta f_\beta}{\lambda_\alpha f_\alpha}\tau_\alpha(z_\beta) \simeq 0.160 \tau_\alpha(z_\beta)\quad.
\label{eq:taubeta}
\end{equation}
Similarly to the \ion{He}{2} Ly$\alpha$ spectra, we concatenated random skewers 
to match the IGM \ion{He}{2} Ly$\beta$ regions of the observed spectra
($3.46 \le z_\beta \le 3.685$ for HE2QS\,J2311$-$1417 and 
$3.56 \le z_\beta \le 3.72$ for HE2QS\,J1630$+$0435). Then we added foreground 
Ly$\alpha$ absorption from random \texttt{Nyx} skewers using the fluctuating 
UV background model by \citet{Davies2017}, resulting in a total optical depth 
$\tau(z_\beta)=\tau_\beta(z_\beta)+\tau_\alpha(z_\mathrm{fg})$.
Figure~\ref{fig:mock_creation} illustrates our procedure for the equivalent 
multiplication of the transmission spectra. The remaining steps are identical 
to the creation of the \ion{He}{2} Ly$\alpha$ mock spectra.

\subsection{Automated Measurement of Transmission Spikes}

We developed a fully automated code that detects and fits all significant 
transmission features with multiple Gaussian profiles. An automated routine is 
crucial for our analysis, because it provides consistent and reproducible 
results for the observed and the fully forward-modeled mock \ion{He}{2} 
absorption spectra, respectively. Due to the strong \ion{He}{2} absorption we 
model multiplicative transmission features with additive Gaussian profiles, 
similar to procedures to model $z\gtrsim 5.5$ \ion{H}{1} transmission spikes 
\citep{Barnett2017,Chardin2018,Gaikwad2020,Yang2020}. While such decompositions 
are entirely empirical, the statistical properties of the detected transmission 
spikes, such as their incidence and their equivalent widths, can be used to 
constrain the ionization state of the IGM. Henceforth, our primary statistic of 
interest is the incidence of transmission spikes, i.e.\ the number of Gaussian 
components in predefined redshift bins.

In order to locate and fit transmission spikes, we must define spectral regions 
that are long enough to reliably fit a Gaussian profile while limiting the 
impact of Poisson noise. First, we calculated the probability $P$ 
(Equation~\ref{eq:prob}) as a $0.36$\,\AA\ (9-pixel) running average to reduce 
the Poisson noise of individual pixels in finding significant transmission.
Then, we considered regions with $n\ge 9$ consecutive pixels with detected 
transmission at $>3\sigma$ significance (running average $P < 0.0014$) for a 
fit of Gaussian profiles. Our choice of $\ge 9$ pixels 
(3--4 resolution elements) is a good compromise between the resolving power and 
the quality (S/N$\lesssim 3$) of our data. The very conservative probability 
limit $P < 0.0014$ ensures that most of the spikes induced by Poisson noise 
will be excluded from the analysis at the expense of the weakest spikes at the 
resolution limit of the data.

In order to find the positions $\lambda _{m}$ of multiple spikes in the 
considered spectral region, we used the Python routine 
\emph{scipy.signal.find\_peaks()} after smoothing the spectra with a Gaussian 
filter $\sigma _\mathrm{f} = 0.1\,\mathrm{\AA}$. With the resulting number of 
components $M$ and their individual positions $\lambda_m$, we fitted their 
heights $A_m$ and widths (Gaussian standard deviation $\sigma_m$) by maximizing 
the Poisson likelihood function (Equation~\ref{eq:poisson-like}) in 
the $n\ge 9$ pixel wide region. The \ion{He}{2} transmission was modeled as the 
sum of the $M$ Gaussian components, such that the modeled counts per pixel in 
Equation~\ref{eq:poisson-like} becomes
\begin{equation}
\label{eq:signal}
S_j = t_j C_j K_{\mathrm{FF},j} E_j\sum _{m=1}^M A_m e^{-(\lambda_m-\lambda_j)^2/(2\sigma_m^2)}\quad.
\end{equation}
Because our fit parameters are not physical but empirical, we refrained from 
computing their statistical errors.

\begin{figure}
	\includegraphics[width=\columnwidth]{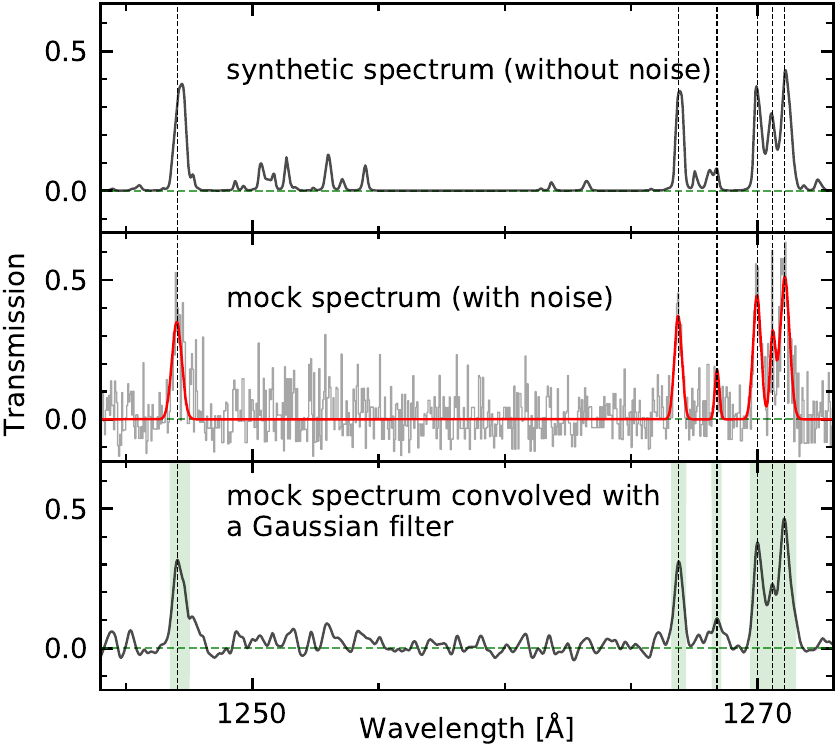}
	\caption{Detected transmission spikes in a \ion{He}{2} Ly$\alpha$ mock spectrum. 
	\emph{Upper panel:} Synthetic spectrum without noise. 
	\emph{Middle panel:} Mock spectrum with Poisson noise (gray) with best-fit 
	Gaussian components (red). 
	\emph{Lower panel:} Noisy mock spectrum convolved with the Gaussian filter 
	with $\sigma_\mathrm{f}=0.1$\,\AA. Green-shaded areas indicate regions of 
	$\ge 9$ consecutive pixels ($\ge 0.36$\,\AA) where the 9-pixel running 
	average of $P$ (Equation~\ref{eq:prob}) is $<0.0014$. The vertical lines 
	indicate the positions of the peak centers found by the Python routine 
	\emph{scipy.signal.find\_peaks()} in the green-shaded areas.}
	\label{fig:spike_finding_method}
\end{figure}

Figure~\ref{fig:spike_finding_method} illustrates the spike finding process for 
a representative \ion{He}{2} Ly$\alpha$ mock spectrum. The comparison between 
the noise-free spectrum (upper panel) and the Gaussian decomposition of the 
mock spectrum (middle panel) shows that the strong transmission peaks are 
found reliably. In the mock spectra $>99$\% (50\%) of the spikes with 
$A_m\ge 0.25$ ($A_m = 0.12$) are recovered. Small spikes, e.g.\ at 
$\lambda\sim 1252$\,\AA\ in Figure~\ref{fig:spike_finding_method}, are often 
lost in the noise as expected. Others, e.g.\ at $\lambda\sim 1268$\,\AA, cannot 
be fitted as precisely as stronger spikes. On average, the algorithm slightly 
overestimates the spike heights due to the asymmetric Poisson noise.
The recovery rate mainly varies with the height of the spikes but not with 
their width. In the end, the recovery rate and the fitting accuracy are not 
particularly relevant, because the same algorithm is applied to the observed 
spectra and to the forward-modeled mock spectra alike.

\section{Results}
\label{sec:results}

\subsection{Fitted Observed \ion{He}{2} Ly$\alpha$ Transmission Spikes}
\label{sec:transmission_in_lya}

\begin{figure*}[t]
	\includegraphics[width=\textwidth]{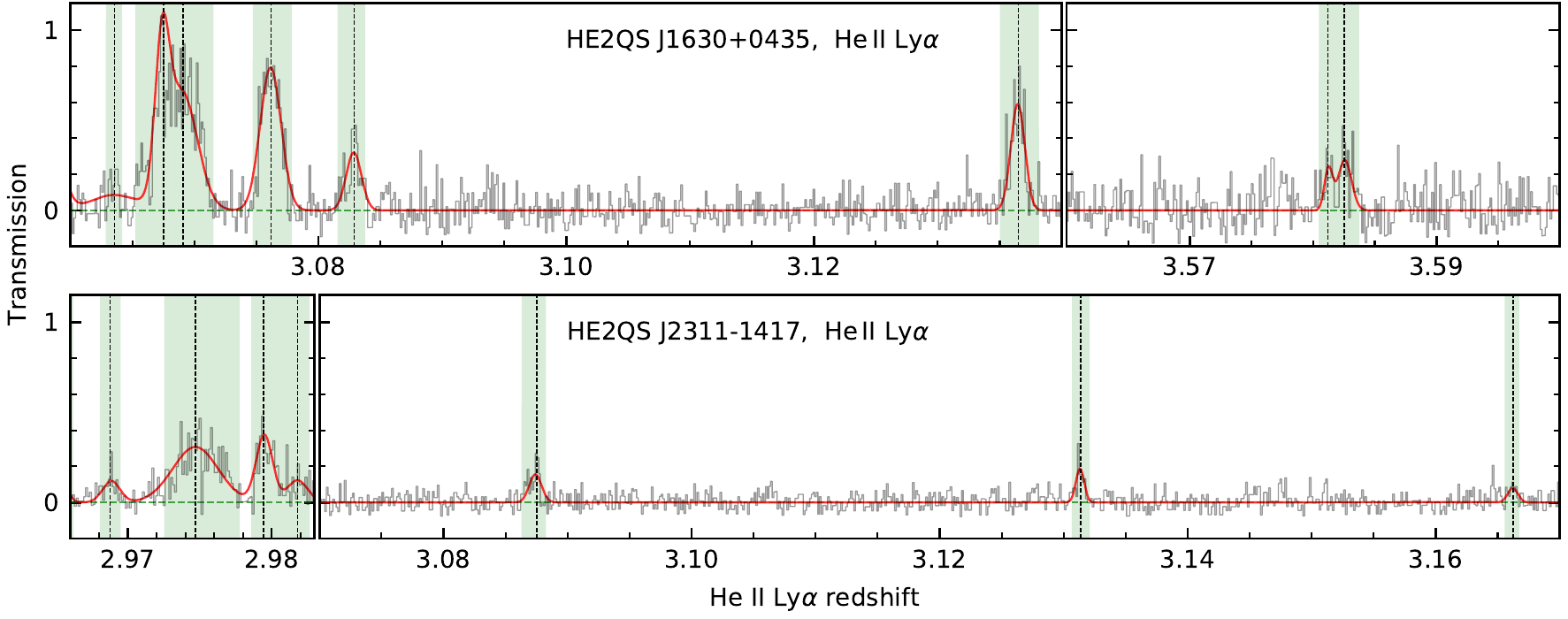}
	\caption{Regions with detected Ly$\alpha$ transmission spikes in the 
	{\it HST}/COS G130M spectra of both sightlines (gray). The green-shaded 
	areas indicate regions with $n \ge $ 9 consecutive pixels with significant 
	transmission ($P \le 0.0014$, Equation~\ref{eq:prob}). Detected peaks 
	(spike locations) are marked with vertical lines. Red curves show the 
	Gaussian decompositions of the sometimes blended transmission features.}
	\label{fig:lya_spikes}
\end{figure*}

\begin{deluxetable}{lllc}
	\tablecaption{Fitting parameters of the \ion{He}{2} Ly$\alpha$ transmission 
	spikes.\label{tab:Lya_spikes}}
	\tablehead{
		\colhead{Object} & \colhead{$z_{\alpha, m}$}  & \colhead{$\sigma _{m}$} & \colhead{$A_{m}$} \\
		\colhead{} & \colhead{} & \colhead{[$\mathrm{km\,s^{-1}}$]} & \colhead{ }}
	\startdata
	HE2QS\,J1630$+$0435 & 3.0636     & 321  	& 0.086 \\	
				        & 3.0675     & 41   	& 0.777 \\
				        & 3.0691     & 92   	& 0.660 \\
				        & 3.0762	 & 62  	    & 0.794 \\
				        & 3.0829     & 45   	& 0.322 \\
				        & 3.1365     & 40    	& 0.593 \\
				        & 3.5812     & 22  	    & 0.231 \\
				        & 3.5825     & 35 	    & 0.280 \\
	HE2QS\,J2311$-$1417	& 2.9688     & 45       & 0.120 \\
                        & 2.9747     & 120      & 0.308 \\
                        & 2.9794     & 43       & 0.374 \\
                        & 2.9818     & 53       & 0.123 \\
                        & 3.0875     & 36       & 0.157 \\
                        & 3.1314     & 24       & 0.189 \\
                        & 3.1662     & 29       & 0.080 \\
    \enddata
\end{deluxetable}

Figure~\ref{fig:lya_spikes} displays the spectral regions in the two sightlines 
that show statistically significant IGM \ion{He}{2} Ly$\alpha$ transmission 
spikes. Most transmission spikes are detected in the narrow redshift range 
$3.06\lesssim z_\alpha\lesssim 3.17$ ($\sim 100$\,cMpc) in both lines of sight.
Their clustered appearance suggests that \ion{He}{2} at $z\sim 3.1$ is highly 
ionized in both sightlines. At higher redshifts we measure two blended spikes 
at $z_\alpha\simeq 3.58$ in the HE2QS\,J1630$+$0435 spectrum. Some of the 
transmission spikes might arise from transverse proximity effects of foreground 
quasars \citep[e.g.,][]{Jakobsen2003, Schmidt2017}, but the largest dedicated 
survey to date \citep{Schmidt2017} did not cover HE2QS\,J2311$-$1417, 
while HE2QS\,J1630$+$0435 requires deeper imaging and follow-up spectroscopy to 
find matching $z_\mathrm{em}>3$ quasars. 
In general, the degeneracies imposed by quasar lifetime and obscuration 
require a statistical analysis \citep{Schmidt2017,Schmidt2018}.
In the HE2QS\,J2311$-$1417 sightline, 
the four spikes at $z_\alpha = 2.97$--$2.98$ are likely IGM Ly$\alpha$ spikes 
rather than Ly$\beta$ spikes of the proximity zone (Figure~\ref{fig:reduced_spectra}), 
because of the lack of correlation with the Ly$\alpha$ proximity zone. However, 
due to the small additional pathlength ($\Delta z_\alpha = 0.017$) between 
\ion{He}{2} Ly$\beta$ and geocoronal Ly$\alpha$ we excluded this region from 
the statistical analysis in Section~\ref{sec:gamma}.

Between these isolated Ly$\alpha$ transmission features both spectra display 
long Gunn-Peterson troughs without significant transmission. The troughs are 
at $3.138 \le z_\alpha \le 3.580$ (383\,cMpc) in the HE2QS\,J1630$+$0435 
spectrum, and at $3.167 \le z_\alpha \le 3.685$ (443\,cMpc) in the 
HE2QS\,J2311$-$1417 spectrum, with $2\sigma$ detection limits of 
$\tau_\mathrm{eff} > 5.82$ and $\tau_\mathrm{eff} > 6.13$, respectively.

Table~\ref{tab:Lya_spikes} lists the fitting parameters of the 
\ion{He}{2} Ly$\alpha$ transmission spikes as individual Gaussian components.  
The combined transmission of the Gaussians at $z_\alpha\simeq 3.065$ in the 
spectrum of HE2QS\,J1630$+$0435 appears unphysical, because it exceeds unity.
However, because our method is entirely empirical, we chose not to limit the 
total transmission. Such effects also appear in the mock spectra and they do not 
affect our results. Likewise, we did not include continuum uncertainty, 
because it does not change the incidence of the spikes.

\subsection{Fitted Observed \ion{He}{2} Ly$\beta$ Transmission Spikes}
\label{sec:tranmission_in_lyb}

\begin{figure}
	\includegraphics[width=\columnwidth]{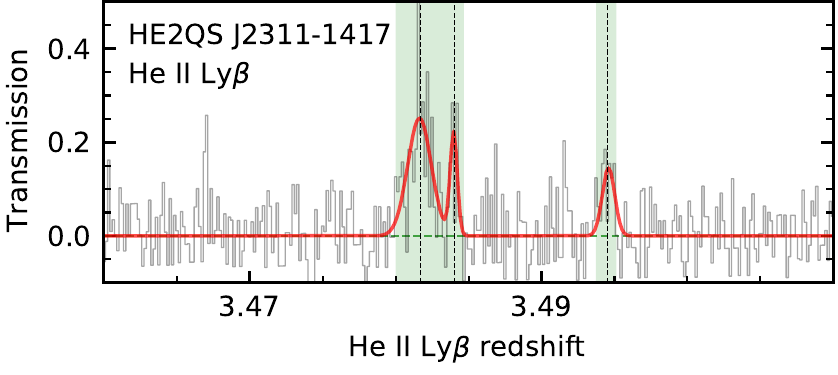}
	\caption{Similar to Figure~\ref{fig:lya_spikes} but for \ion{He}{2} Ly$\beta$.}
	\label{fig:Lyb_model}
\end{figure}

\begin{deluxetable}{lllc}
	\tablecaption{Fitting parameters of the \ion{He}{2} Ly$\beta$ 
	transmission spikes.\label{tab:Lyb_spikes}}
	\tablehead{
		\colhead{Object} & \colhead{$z_{\beta, m}$}  & \colhead{$\sigma _{m}$} & \colhead{$A_{m}$} \\
		\colhead{ } & \colhead{ } & \colhead{[$\mathrm{km\,s^{-1}}$]} & \colhead{ }}
	\startdata
	HE2QS\,J2311$-$1417	    & 3.4817     & 54    & 0.252  \\
	                        & 3.4840     & 16    & 0.220  \\
	                        & 3.4945     & 28    & 0.146  \\
	\enddata
\end{deluxetable}

\begin{figure*}[t]
	\includegraphics[width=\textwidth]{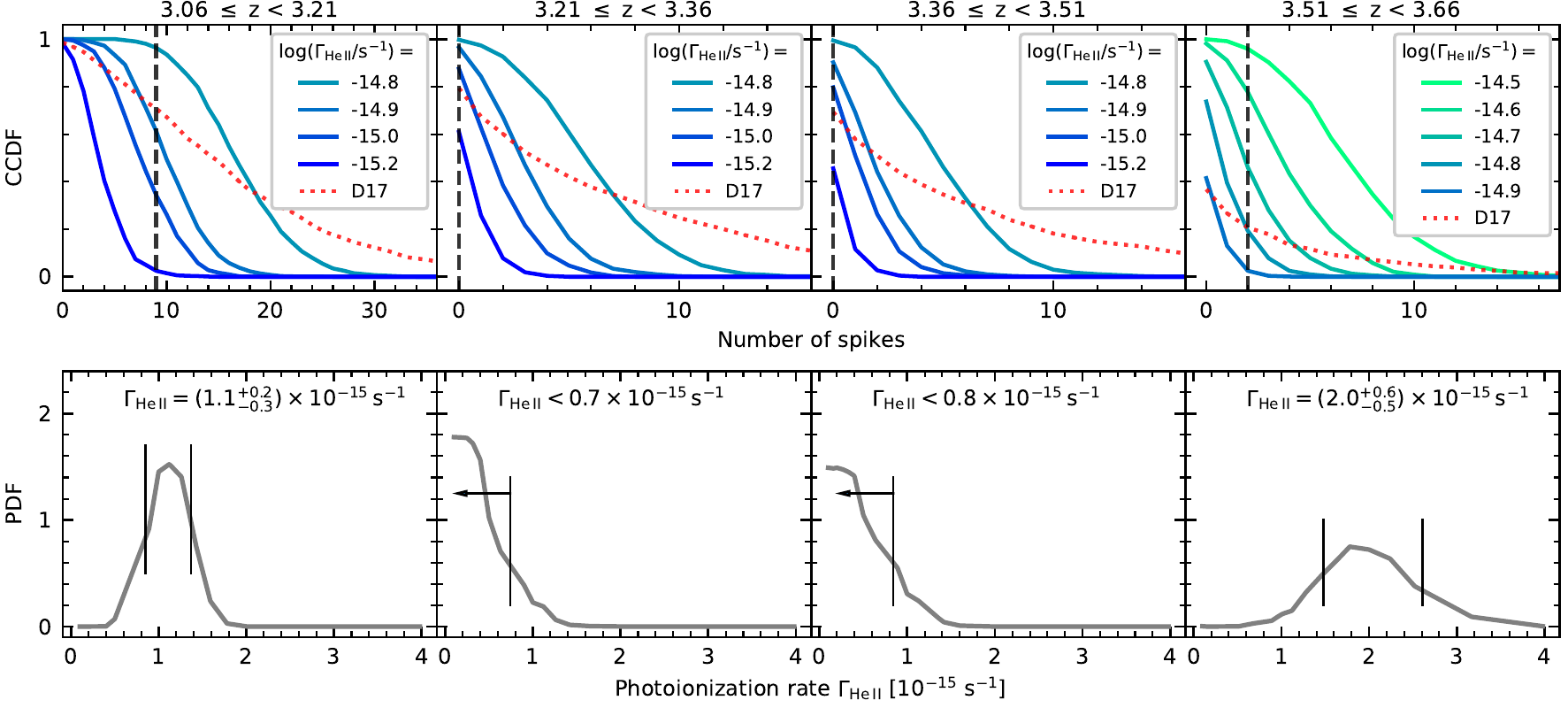}
	\caption{Top: CCDFs of the incidence of \ion{He}{2} Ly$\alpha$ spikes for 
	different uniform photoionization rates in $\Delta z = 0.15$ derived from 
	realistic mock COS spectra for our two sightlines. Although the CCDFs are 
	discrete, they are plotted as non-discrete distributions for illustration 
	purposes. The vertical dashed lines show the number of detected spikes in 
	our observed spectra in the corresponding redshift bin. The red 
	dotted lines indicate CCDFs with the fluctuating UV background model 
	from \citet{Davies2017} shown here for comparison. Bottom: Normalized 
	posterior probability density functions (PDFs) of $\Gamma_{\mathrm{He\,II}}$ 
	given the measured number of spikes (Equation \ref{eq:pdf_alpha}). 
	The posterior median yields our inferred $\Gamma_\mathrm{HeII}$ value 
	(labeled), with the 16th--84th percentile range (vertical lines) 
	as $1\sigma$ uncertainty. 
	}
	\label{fig:cumul_Lya_n}
\end{figure*}

The HE2QS\,J2311$-$1417 spectrum contains three significant 
\ion{He}{2} Ly$\beta$ transmission spikes at $z_\beta\sim 3.49$, presented 
in Figure~\ref{fig:Lyb_model} and tabulated in Table~\ref{tab:Lyb_spikes}.
The strongest of these was already foreshadowed by a single high-transmission 
pixel in the G140L spectrum, which is as unreliable as most individual 
Ly$\alpha$ transmission pixels (Figures~\ref{fig:opt_depth_j1630} and 
\ref{fig:opt_depth_j2311}). Here we confirm this suggestion with superior data.
This is the first time that significant \ion{He}{2} Ly$\beta$ transmission is 
detected and resolved at $z_\beta>3.4$, suggesting the presence of highly 
ionized patches in the $z>3.4$ IGM that are saturated in \ion{He}{2} Ly$\alpha$ 
at \ion{He}{2} fractions of a few percent \citep{McQuinn2009}.
Previous studies considered lower redshifts, and focused on the 
\ion{He}{2} Ly$\beta$ effective optical depth to reach higher sensitivity in 
\ion{He}{2} Ly$\alpha$ \citep{Zheng2004,Syphers2011}. The results were plagued 
by low data quality and systematics due to the required modeling of 
\ion{He}{2} Ly$\alpha$ foreground absorption \citep{Syphers2011}.
Here, we capitalize on spectral resolution, data quality, redshift coverage and 
detailed mock spectra to better constrain the ionization state of the IGM.
Finally, while there are hints that also the HE2QS\,J1630$+$0435 sightline 
shows IGM \ion{He}{2} Ly$\beta$ transmission (Appendix~\ref{sec:geocor_decont}), 
we cannot definitely prove its origin, so we exclude it from further analysis 
(see Appendix~\ref{sec:likely_lyb_spikes} for the alternative assertion).

We emphasize that both sightlines had been selected for observation solely 
because of their FUV brightness and thus, our very modest sample is unbiased 
with respect to the spike incidence. Thus, the very fact that we detect 
\ion{He}{2} Ly$\beta$ transmission at $z_\beta\simeq 3.49$ in the 
HE2QS\,J2311$-$1417 sightline despite foreground contamination, together with 
definite Ly$\alpha$ transmission at $z_\alpha\simeq 3.58$ and possible Ly$\beta$ 
transmission at $z_\beta\simeq 3.68$ toward HE2QS\,J1630$+$0435, supports 
suggestions of patchy \ion{He}{2} reionization (i.e.\ a fluctuating 
UV background) at $z > 3.4$ \citepalias{Worseck2019}.

\subsection{Implications for the \ion{He}{2} Photoionization Rate}
\label{sec:gamma}

Evidently, a higher \ion{He}{2} photoionization rate increases the number and 
amplitude of \ion{He}{2} Ly$\alpha$ transmission features (Figure~\ref{fig:growing_forest}).
In order to study this dependence in more quantitative terms, we constructed 
complementary cumulative distribution functions (CCDFs) from the 
probability mass functions of the incidence of 
spikes in our mock spectra. The CCDF$(n)$ is the fraction of mock spectra 
with more than $n$ detected spikes. For every grid point in 
$\Gamma _{\mathrm{HeII}}$, we computed the CCDF of Ly$\alpha$ transmission 
spikes from mock spectra of both sightlines in $\Delta z = 0.15$ redshift bins.
By using $\Delta z = 0.15$ bins (combined pathlength  $\sim 260\mathrm{\,cMpc}$), 
we decrease the impact of density fluctuations on the measurement while tracking 
the redshift evolution of the photoionization rate. The combination of the 
individual sightlines is justified due to their similar spike recovery rates.

Figure~\ref{fig:cumul_Lya_n} shows the resulting CCDFs for representative 
photoionization rate grid values. Although each CCDF spans a wide range due to 
IGM density fluctuations, the CCDFs monotonically shift to higher numbers of 
Ly$\alpha$ spikes with increasing $\Gamma_{\mathrm{HeII}}$. Therefore, while 
our spike decomposition is empirical, we are able to infer a physical parameter, 
the characteristic \ion{He}{2} photoionization rate from the measured spike 
incidence. Thanks to our forward-modeling, the CCDFs include the instrumental 
effects and data quality limitations. 
For comparison, we also computed 
CCDFs for the fluctuating UV background from \citet{Davies2017} which are consistent with 
our measured spike incidence. For example at $z \sim 3.1$, in 40\,\% of the mock data
there are between 5 and 13 spikes. Furthermore, the non-detection at $z = 3.2$--$3.5$ 
is very common since $\sim$25\,\% of the mock data show no spikes.

In order to infer the \ion{He}{2} photoionization rates, we have to assume that 
our two sightlines are representative of the IGM. We determined 
$\Gamma_\mathrm{HeII}$ from the number of detected \ion{He}{2} Ly$\alpha$ 
transmission spikes in $\Delta z = 0.15$ bins. For this, we constructed 
posterior probability distributions $p(\Gamma _{\mathrm{He\,II}}|n_{\alpha})$ for 
the number of detected \ion{He}{2} Ly$\alpha$ spikes $n_{\alpha}$ 
(Figure~\ref{fig:cumul_Lya_n}) by using Bayes' theorem
\begin{equation}
\label{eq:pdf_alpha}
p(\Gamma _{\mathrm{He\,II}}|n_{\alpha}) \propto L(n_{\alpha}|\Gamma _{\mathrm{He\,II}})p(\Gamma _{\mathrm{He\,II}})
\end{equation}
with a uniform prior $p(\Gamma _{\mathrm{He\,II}})$ on a refined 
grid $\Delta[\log (\Gamma _{\mathrm{HeII}}/\mathrm{s^{-1}})] = 0.05$ 
to adequately sample the posterior and the likelihood 
$L(n_{\alpha}|\Gamma _{\mathrm{He\,II}})$ derived from the probability mass 
functions of the spike incidence. The posterior was normalized to unit integral.
We quote the median value of the posterior distribution as our $\Gamma_{\mathrm{He\,II}}$ 
measurement, and the equal-tailed 84th--16th percentile range as its $1\sigma$ 
uncertainty. The redshift bins without detected transmission spikes 
yield $1\sigma$ upper limits from the 84th percentile of the posterior.  

We detect nine spikes at $3.06 \le z < 3.21$ and two spikes at $ 3.51 \le z < 3.66$, 
resulting in two measurements of the \ion{He}{2} photoionization rate 
(Table~\ref{tab:photion_results}). In the other two redshift bins we derive only 
upper limits due to the lack of spikes. Repeating the above analysis for the 
total equivalent width of the detected spikes instead of their incidence 
resulted in similar $\Gamma_\mathrm{He\,II}$ values within the broad 
confidence intervals. Somewhat larger redshift bins ($\Delta z = 0.2$) do not 
significantly change our results either.

\begin{deluxetable}{lccl}
	\tablecaption{Median $\Gamma _{\mathrm{He\,II}}$ inferred from the number 
	of \ion{He}{2} Ly$\alpha$ or Ly$\beta$ transmission spikes in redshift bins 
	$\Delta z$. \label{tab:photion_results}}
	\tablehead{\colhead{Object} & \colhead{$\Delta z$} & \colhead{$\mathrm{\Gamma _{\mathrm{He\,II}} (10^{-15}\,s^{-1})}$} & \colhead{Transition}}
	\startdata
    Both                &  $3.06$--$3.21$ & $1.1^{+0.2}_{-0.3}$ & Ly$\alpha$\\
                        &  $3.21$--$3.36$ & $< 0.7$             & Ly$\alpha$\\
                        &  $3.36$--$3.51$ & $< 0.8$             & Ly$\alpha$\\
                        &  $3.51$--$3.66$ & $2.0^{+0.6}_{-0.5}$ & Ly$\alpha$\\
    \hline
    HE2QS\,J2311$-$1417 &  $3.46$--$3.685$  & $1.1^{+0.9}_{-0.4}$ & Ly$\beta$\\
                        &                   & $0.9\pm0.3$       & Ly$\beta$ and Ly$\alpha$\\
    HE2QS\,J1630$+$0435 &  $3.56$--$3.72$ & $< 1.9 $            & Ly$\beta$\\
                        &                   & $2.3^{+0.7}_{-0.6}$       & Ly$\beta$ and Ly$\alpha$\\
	\enddata
\end{deluxetable}

\begin{figure}
	\centering
	\includegraphics[width=\columnwidth]{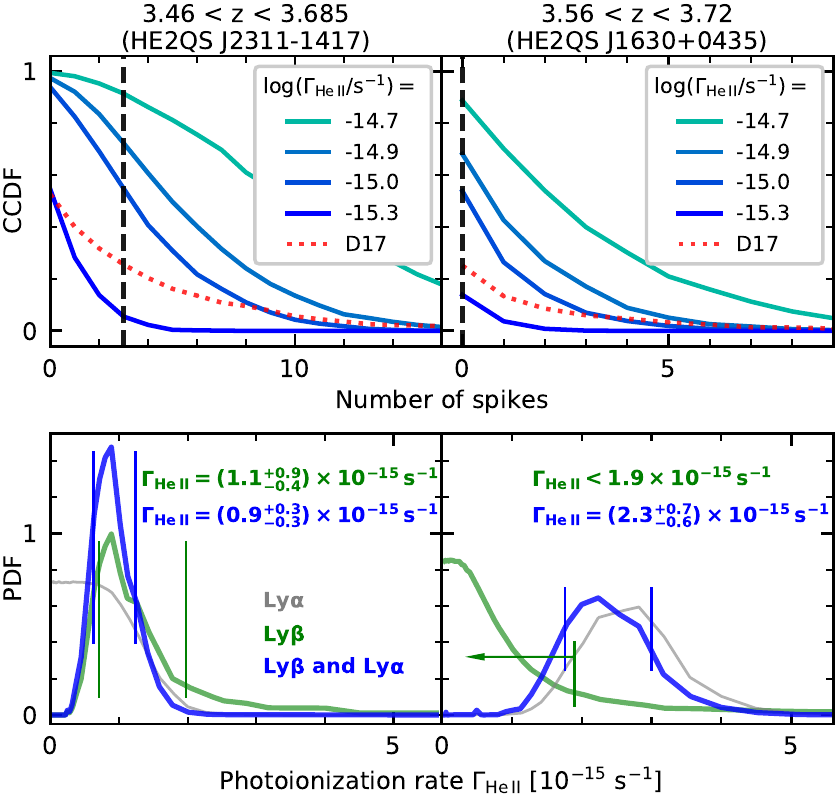}
	\caption{Similar to Figure~\ref{fig:cumul_Lya_n} but for 
	\ion{He}{2} Ly$\beta$ and separately for each sightline.
	The bottom panel shows the PDFs inferred from 
	Ly$\alpha$ (gray), Ly$\beta$ (green) and from both transitions combined (blue).
	\label{fig:cumul_Lyb_n}}
\end{figure}

In contrast to \ion{He}{2} Ly$\alpha$, the spectral regions covering 
\ion{He}{2} Ly$\beta$ in both sightlines have only a short overlap of 
$\Delta z \sim 0.12$. Thus, we analyzed the sightlines independently 
(Figure \ref{fig:cumul_Lyb_n}). Similar to \ion{He}{2} Ly$\alpha$, the 
fluctuating UV background model \citep{Davies2017} is consistent with the 
measured \ion{He}{2} Ly$\beta$ spike incidence, e.g.\ in 35\,\% of 
sightlines at $z=3.46$--$3.685$ there are between 1 and 5 spikes.
The posteriors inferred just from the detected \ion{He}{2} Ly$\beta$ 
spikes $n_{\beta}$ are highly asymmetric and wider in comparison to those 
of \ion{He}{2} Ly$\alpha$, mostly due to the required \ion{He}{2} Ly$\alpha$ 
foreground modeling using the fluctuating UV background. 
Since the Ly$\beta$ redshift range of each sightline has simultaneous 
Ly$\alpha$ coverage we used a joint likelihood for detecting $n_{\alpha}$ 
and $n_{\beta}$ spikes to construct the combined posterior
\begin{equation}
\label{eq:pdf_beta}
p(\Gamma _{\mathrm{He\,II}}|n_{\alpha},  n_{\beta}) \propto L(n_{\alpha}|\Gamma _{\mathrm{He\,II}})L(n_{\beta}|\Gamma _{\mathrm{He\,II}})p(\Gamma _{\mathrm{He\,II}})
\end{equation}
The spike incidences $n_{\alpha}$ and $n_{\beta}$ can be 
considered independent due to the overlapping foreground 
\ion{He}{2} Ly$\alpha$ absorption. By using the combined posterior we 
obtain a more precise $\Gamma _{\mathrm{HeII}}$ constraint toward HE2QS\,J2311$-$1417 
($\Gamma _{\mathrm{HeII}} = (0.9\pm0.3)\times10^{-15}\mathrm{\,s^{-1}}$ 
for $n_{\alpha} = 0$ and $n_{\beta} = 3$ at $z=3.46$--$3.685$)
and an actual measurement toward HE2QS\,J1630$+$0435 instead of the upper limit 
($\Gamma _{\mathrm{HeII}} = (2.3^{+0.7}_{-0.6})\times10^{-15}\mathrm{\,s^{-1}}$ 
for $n_{\alpha} = 2$ and $n_{\beta} = 0$ at $z=3.56$--$3.72$).
The values are consistent with those obtained from Ly$\alpha$ at $2\sigma$ (Table \ref{tab:photion_results}).

\begin{figure}
	\centering
	\includegraphics[width=\columnwidth]{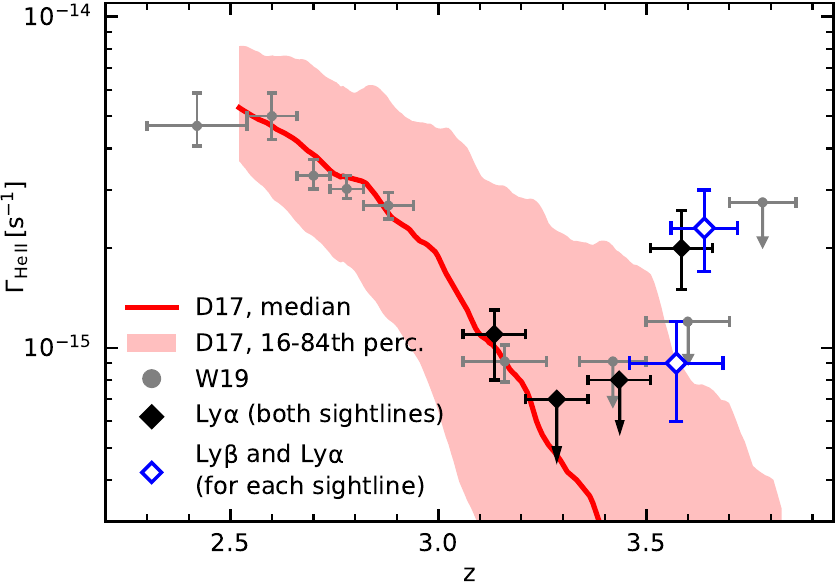}
	\caption{
	Redshift evolution of the \ion{He}{2} photoionization rate inferred from 
	the incidence of transmission spikes (diamonds) and from the effective 
	optical depth \citepalias[circles,][]{Worseck2019}. Error bars show the 
	16th--84th percentile range derived from the $\Gamma _{\mathrm{He\,II}}$ 
	posterior given the measured number of spikes. Upper limits are inferred 
	from 84th percentile of the $\Gamma _{\mathrm{He\,II}}$ posterior. 
	The red curve shows the median $\Gamma_\mathrm{He\,II}(z)$ and its 
	16th--84th percentile range (red-shaded) from 3D semianalytic 
	calculations \citep{Davies2017}.}
	\label{fig:photion_n}
\end{figure}

The measured $\Gamma_{\mathrm{He\,II}}$ values are plotted in 
Figure~\ref{fig:photion_n} along with the results from \citetalias{Worseck2019} 
and the prediction from the fluctuating UV background model by \citet{Davies2017}.
At $z < 3.5$, our new measurements are consistent with the model and 
$\Gamma_{\mathrm{He\,II}}$ derived from $\tau_{\mathrm{eff}}$ measurements 
\citepalias{Worseck2019}. At $z > 3.5$ the measurements are higher than expected 
because of the detected \ion{He}{2} Ly$\alpha$ and Ly$\beta$ spikes along both 
lines of sight. Nevertheless, they are roughly consistent with the expected 
$\Gamma_{\mathrm{HeII}}$ fluctuations in the \citet{Davies2017} model. 
This consistency is further supported with the D17 CCDFs shown in 
Figures \ref{fig:cumul_Lya_n} and \ref{fig:cumul_Lyb_n}. A measurement of a 
representative median $\Gamma_{\mathrm{HeII}}$ for the IGM as a whole 
requires a substantially larger sample.

\section{Summary}
\label{sec:summary}

We have presented and analyzed new high-resolution ($R \sim 14,000$) 
{\it HST}/COS spectra of the two UV-brightest \ion{He}{2}-transparent quasars 
at $z_\mathrm{em} > 3.5$ (Figure~\ref{fig:reduced_spectra}). The high resolution 
and data quality (S/N $\sim 3$ per $0.04\mathrm{\,\AA}$ pixel) for the first 
time enables us to study narrow resolved (FWHM $\gtrsim 50\mathrm{\,km\,s^{-1}}$) 
\ion{He}{2} transmission spikes at $z > 3.4$ originating in highly ionized 
underdense regions in the IGM. We have analyzed the incidence of 
\ion{He}{2} Ly$\alpha$ and Ly$\beta$ transmission spikes in both spectra, using 
a new fully automatic routine for our Poisson-limited data that enables a 
one-to-one comparison to forward-modeled spectra from numerical simulations.
Our main results are the following:

\begin{enumerate}
    \item We confirm the most prominent $z>3.06$ Ly$\alpha$ spikes from previous 
    low-resolution data \citepalias{Worseck2019} and measure their properties 
    (Figure~\ref{fig:lya_spikes}). The vast majority of the 
    \ion{He}{2} Ly$\alpha$ spikes (13/15) reside at $z < 3.2$, indicating mostly 
    ionized intergalactic helium in both sightlines at these redshifts, in 
    agreement with the $\tau _{\mathrm{eff}}$ measurements from the larger 
    low-resolution sample \citepalias{ Worseck2019}. However in one sightline 
    (HE2QS\,J1630$+$0435), two \ion{He}{2} Ly$\alpha$ spikes at $z\simeq3.58$ 
    adjacent to a $\sim 383$\,cMpc Gunn-Peterson trough suggest the occasional 
    presence of highly ionized underdense IGM regions also at higher redshifts. 

    \item We report the first detection of a group of three resolved 
    \ion{He}{2} Ly$\beta$ transmission spikes at $z \sim 3.49$ toward 
    HE2QS\,J2311$-$1417, where \ion{He}{2} Ly$\alpha$ is fully saturated. This 
    further supports our inferences from \ion{He}{2} Ly$\alpha$ transmission in 
    the other sightline at similar redshifts. 
    
    \item We inferred the IGM \ion{He}{2} photoionization rates at 
    $3.1 \lesssim z \lesssim 3.6$ by comparing our measured spike incidence to 
    predictions from forward-modeled mock spectra from a 
    $(146\,\mathrm{cMpc})^{3}$ hydrodynamical simulation. Despite the 
    limitation of our small sample and our assumption of a spatially 
    uniform UV background, our inferred 
    $\Gamma _{\mathrm{HeII}} \sim 10^{-15}\mathrm{\,s^{-1}}$ at $z \simeq 3.1$ 
    is comparable to previous measurements and models. At higher redshifts, the 
    transmission spikes detected in both sightlines result in higher 
    $\Gamma_{\mathrm{HeII}}$. However, although these values may not be 
    representative of the average $\Gamma_{\mathrm{HeII}}$ in the IGM, they suggest a 
    fluctuating UV background at the end of the \ion{He}{2} reionization 
    epoch \citep{Davies2017}.

\end{enumerate}

This is the first time that small-scale structure in high-redshift \ion{He}{2} 
absorption has been detected and consistently analyzed, similarly to \ion{H}{1} 
studies at $z > 5.5$ \citep{Chardin2018, Garaldi2019, Gaikwad2020, Yang2020}.
We conclude that \ion{He}{2} Ly$\alpha$ and Ly$\beta$ transmission spike 
statistics are promising tools to probe the tail end of the 
\ion{He}{2} reionization epoch. Further progress will require improved 
large-volume high-resolution hydrodynamical simulations capable of resolving 
underdense voids that dominate the \ion{He}{2} absorption. While including full 
radiative transfer in these simulations may still be infeasible, approximate 
methods successfully capture the large-scale fluctuations of the 
\ion{He}{2}-ionizing background that give rise to the observed transmission 
spikes \citep{Davies2017}. Observationally, our very modest sample of two 
$z>3.5$ sightlines may be increased by high-resolution \textit{HST}/COS 
follow-up of recently discovered FUV-bright quasars (e.g., \textit{HST} 
program 16317, PI Worseck). 
However, at $z\ga 3.2$ where a study of patchy \ion{He}{2} absorption is 
worthwhile, there are currently less than ten quasars bright enough for a 
high-resolution \textit{HST} spectroscopy, four of which have been observed until now. 
Moreover, with our simulations and the \citet{Davies2017} model we predict 
that $\sim 30$\% of the \ion{He}{2} Ly$\alpha$ transmission spikes have 
$\mathrm{FWHM}=10$--$20\mathrm{\,km\,s^{-1}}$. These will be securely detected and 
resolved only with FUV-sensitive spectrographs onboard future large space 
telescopes \citep[e.g., LUVOIR-A/LUMOS, ][]{LUVOIR2019} that will gather 
routinely $R\sim 30,000$ spectroscopy of more abundant faint (FUV$\sim 23$) 
\ion{He}{2}-transparent quasars.

\acknowledgments

We thank the anonymous referee for very helpful comments. 
This work was funded by Bundesministerium f\"ur Wirtschaft und Energie in the 
framework of the Verbundforschung of the Deutsches Zentrum f\"ur Luft- und 
Raumfahrt (DLR, grant 50 OR 1813). Support for program GO~15356 was provided by 
NASA through a grant from the Space Telescope Science Institute, which is 
operated by the Association of Universities for Research in Astronomy, Inc., 
under NASA contract NAS5-26555.

This research made use of Astropy, a community-developed core Python
package for Astronomy \citep{Astropy2013, Astropy2018}. 

\facility{\textit{HST} (COS)}

\software{astropy \citep{Astropy2013,Astropy2018},
          SciPy \citep{Scipy2020},
          numpy \citep{vanderWalt2011},
          matplotlib \citep{Hunter2007}
          }

\appendix\twocolumngrid

\section{FaintCOS -- An Improved \textit{HST}/COS Reduction Pipeline for Faint Objects}\label{sec:faintcos}

It is known that the standard \textit{HST}/COS reduction pipeline 
\texttt{CALCOS} miscalculates the COS FUV detector dark current due to the 
inhomogeneously degrading detector sensitivity \citep{Syphers2012,Worseck2016}.
Because the accurate estimation of the background is essential for the analysis 
\textit{HST}/COS spectra of faint objects 
($f_\lambda\lesssim 10^{-16} \mathrm{\,erg\,cm^{-2}\,s^{-1}\,\AA ^{-1}}$), we 
developed an improved reduction pipeline \texttt{FaintCOS},
a publicly available Python code for a fast and science-grade reduction of 
COS FUV spectra. Building on \texttt{CALCOS}, \texttt{FaintCOS} offers 
(1) a streamlined customization of \texttt{CALCOS} reduction parameters, 
(2) accurate dark current estimation with post-processed dark frames \citep{Worseck2016}, 
and (3) science-grade co-addition of sub-exposures across different COS 
wavelength settings and \textit{HST} data sets that preserves COS Poisson counts, 
and calculates correct confidence intervals accounting for the 
background \citep{FeldmanCousins1998}.

\subsection{Optimized Detector Pulse Heights and Point Source Extraction Windows}

\texttt{FaintCOS} reduces the impact of the dark current onto the extracted 
spectrum by employing custom pulse height amplitude (PHA) limits that reflect 
the state of the COS FUV detector at the time of observation, and by using 
narrow boxcar source extraction apertures that preserve the spectrophotometry 
of point sources. Boxcar extraction is adequate, because the more recent 
two-zone extraction algorithm (a boxcar with variable width) implemented in 
\texttt{CALCOS} is not applicable to dark frames.

\begin{figure}[t]
	\includegraphics[width=\columnwidth]{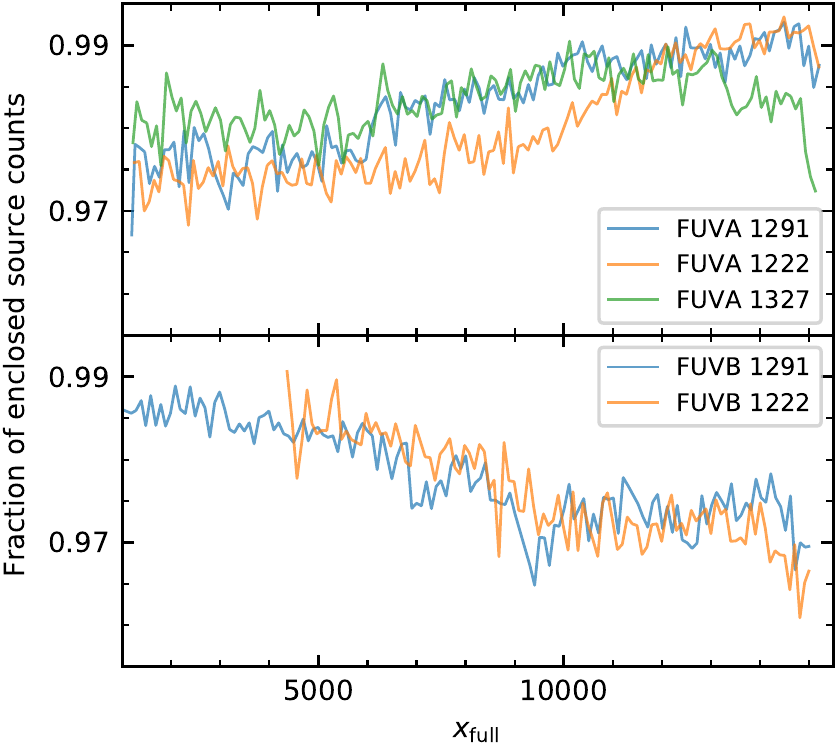}
	\caption{Fraction of source counts in the custom rectangular boxcar 
	extraction apertures in the dispersion direction on COS detector 
	segments A and B for our used COS wavelength settings.
	}
	\label{fig:frac_of_counts}
\end{figure}

The boxcar source extraction aperture was defined individually for every COS 
detector lifetime position, FUV grating, central wavelength, and COS detector 
segment using the corrected event lists of standard stars\footnote{WD~0308$-$565 
(Programs 12426, 12806, 13353, 13932, 14910, 15367, 15384, 15458, 15535),  
WD~0947$+$857 and WD~1057$+$719 (Program 11897).} recorded at high S/N.
We analyzed their cross-dispersion profiles in $1\,\mathrm{\AA}\,$ bins, and 
determined upper and lower corrected cross-dispersion detector coordinates 
$y_\mathrm{full}$, which enclose 95\% of the source counts in the bin.
The total lowest and highest integer $y_\mathrm{full}$ of all bins determined 
the rectangular extraction aperture. We verified by eye that the apertures are 
sensible, disregarding the detector edges. In this way, $>95$\% of the light of 
a point source is included in the final apertures (Figure~\ref{fig:frac_of_counts}), 
while reducing the dark current by $\sim 40$\% compared to the default 
\texttt{CALCOS} boxcar extraction windows. We found $< 10$\% deviations of the 
trace position for different visits with the same setup, which can be explained 
by statistical fluctuations, round-off errors, or small pointing inaccuracies.
While our optimized apertures are reasonable for point sources, the user can 
adjust them to the science objective.

\texttt{FaintCOS} also uses more rigorous PHA limits, as described 
by \citet{SyphersShull2013} and \citet{Worseck2016}. Due to detector gain sag 
and voltage increases the PHA limits of source counts evolve with time and 
should be adjusted accordingly in \texttt{FaintCOS}. The PHA limits can be 
estimated by examining the PHA distribution of external counts 
(source and geocoronal emission) in the extraction window \citep{Worseck2016}.
Lowering the PHA range from the standard $2 \leqslant \mathrm{PHA} \leqslant 23$ 
to the actual range of our science data $2 \leqslant \mathrm{PHA} \leqslant 12$ 
reduced the dark current by another $\sim 35$\% while retaining $>99$\% of 
the source counts.

\subsection{Accurate Estimation of the COS Dark Current}
\label{sec:dc_estimation}

The COS FUV dark current is estimated from dark monitoring data taken within a 
user-specified time window (default is two months) around the observation date 
of the science data to capture the state of the FUV detector in terms of 
(1) gain sag due to illumination and (2) dark current variation due to varying 
environmental conditions \citep{Worseck2016}.

\begin{deluxetable*}{lccl}[t]
	\tablecaption{FITS table columns of the final co-added spectrum.
	\label{tab:coadd_result}}
	\tablehead{
		\colhead{Column Name} & \colhead{Data Type} & \colhead{Units} & \colhead{Description} }
	\tabletypesize{\small}
	\startdata
	WAVELENGTH 				& 	float & $\mathrm{\AA} $								&	Wavelength	\\
	FLUX	 				& 	float & $\mathrm{erg\,s^{-1}\,cm^{-2}\,\AA ^{-1}} $	&	Flux density $f_{\lambda}$ (Equation~\ref{eq:flux})	\\
	FLUX\_ERR\_UP 			& 	float & $\mathrm{erg\,s^{-1}\,cm^{-2}\,\AA ^{-1}} $	&	Upper uncertainty for the flux density\tablenotemark{a} \\
	FLUX\_ERR\_DOWN 		& 	float & $\mathrm{erg\,s^{-1}\,cm^{-2}\,\AA ^{-1}} $ &	Lower uncertainty for the flux density\tablenotemark{a} 	\\
	GCOUNTS 				&   integer & $\mathrm{counts} $							&	Gross counts $N$	\\
	BACKGROUND 				& 	float & $\mathrm{counts} $							&	Total background $ B = B_\mathrm{dark} + B_\mathrm{Ly\alpha}$  	\\
	BKG\_ERR\_UP			& 	float & $\mathrm{counts} $							&	Upper background error $\sigma_{B,\mathrm{up}} = \sqrt{\sigma_\mathrm{dark}^2 + \sigma_\mathrm{Ly\alpha,up}^2}$ 	\\
	BKG\_ERR\_DOWN			& 	float & $\mathrm{counts} $							&	Lower background error $\sigma_{B,\mathrm{down}} = \sqrt{\sigma_\mathrm{dark}^2 + \sigma_\mathrm{Ly\alpha,down}^2}$ 	\\
	DARK\_CURRENT			& 	float & $\mathrm{counts} $							&	Estimated dark current $B_{\mathrm{dark}}$	\\
	DARK\_CURRENT\_ERR		& 	float & $\mathrm{counts} $							&	Dark current error $\sigma_\mathrm{dark}$ 	\\
	EXPTIME 				& 	float & $\mathrm{seconds} $							&	Exposure time $t_{\mathrm{exp}}$	\\
	DQ		 				& integer & 											&	Data quality flag\tablenotemark{b} 	\\
	CALIB	 				& 	float & $\mathrm{counts\,cm^{2}\,\AA \, erg^{-1}} $	&	Flux calibration curve $C$	\\
	FLAT\_CORR 				& 	float & 											&	Flatfield correction factor $K_{\mathrm{FF}}$	\\
	LYA\_SCATTER			& 	float & $\mathrm{counts} $							&	Scattered geocoronal Ly$\alpha$ emission $B_{\mathrm{Ly\alpha}}$  	\\
	LYA\_SCATTER\_ERR\_UP	& 	float & $\mathrm{counts} $							&	Upper error $\sigma _{\mathrm{Ly\alpha, up}}$ of $B_{\mathrm{Ly\alpha}}$ 	\\
	LYA\_SCATTER\_ERR\_DOWN	& 	float & $\mathrm{counts} $							&	Lower error $\sigma _{\mathrm{Ly\alpha, down}}$ of $B_{\mathrm{Ly\alpha}}$ 	\\
	\enddata
	\tablenotetext{a}{The sum of lower and upper uncertainty give a double-sided 
	Poisson $1\sigma$ uncertainty corresponding to an $68.26$\% confidence level. 
	Methods are either frequentist \citep{FeldmanCousins1998} augmented by 
	Monte Carlo simulations for $N<B$, or Bayesian \citep[][shortest $68.26$\% 
	confidence interval around posterior maximum]{Kraft1991}.
	}
	\tablenotetext{b}{Lowest DQ value of all co-added counts in the wavelength 
	bin adopted from \texttt{CALCOS}. DQ$=0$ indicates no anomalies.
	}
	
\end{deluxetable*}

To account for the spatial and temporal variations in the FUV detector dark 
current, \texttt{FaintCOS} uses the PHA distribution in unilluminated parts of 
the detector that is sensitive to the environmental conditions (thermospheric 
density, cosmic ray hit rate) that vary with solar activity and geomagnetic 
latitude \citep{Worseck2016}. A subset of dark frames taken in similar 
conditions as the science exposure is selected by comparing the respective 
cumulative PHA distributions obtained from two predefined calibration windows 
above and below the science extraction aperture. Regions around geocoronal 
emission lines are excluded to avoid a change of the PHA distribution by 
scattered light. We selected dark frames whose normalized cumulative PHA 
distribution have a maximum absolute difference $D<0.03$ with respect to the 
PHA distribution of the science exposure. The threshold $D$ was increased if 
fewer than 5 dark frames were selected. Both $D$ and the minimum number of 
dark frames can be set by the user according to the variability of solar 
activity and the availability of dark frames.

The final sample of dark frames is stacked and shifted to the focal-plane 
offset position of the science exposure. In the stack, the dark current is 
extracted from the science aperture, and smoothed with a running average. We 
chose a 500-pixel wide window that results in a robust estimate of the dark 
current including its spatial structure (Appendix~\ref{sec:dc_validation_test}).
Pixels with data quality flags $>0$ are excluded from the running average due 
to spatial or temporal detector issues. The systematic error of the averaged 
dark current is estimated with a Poisson standard deviation 
$\sqrt{N_\mathrm{dark}}$, where $N_\mathrm{dark}$ is the number of counts in 
the averaging window. The smoothed dark current and its error are scaled to the 
science exposure using the total number of counts in the calibration windows, 
yielding the final dark current estimate $B_\mathrm{dark}$ and its 
error $\sigma_\mathrm{dark}$.

\begin{figure*}
	\centering
	\includegraphics[width=\textwidth]{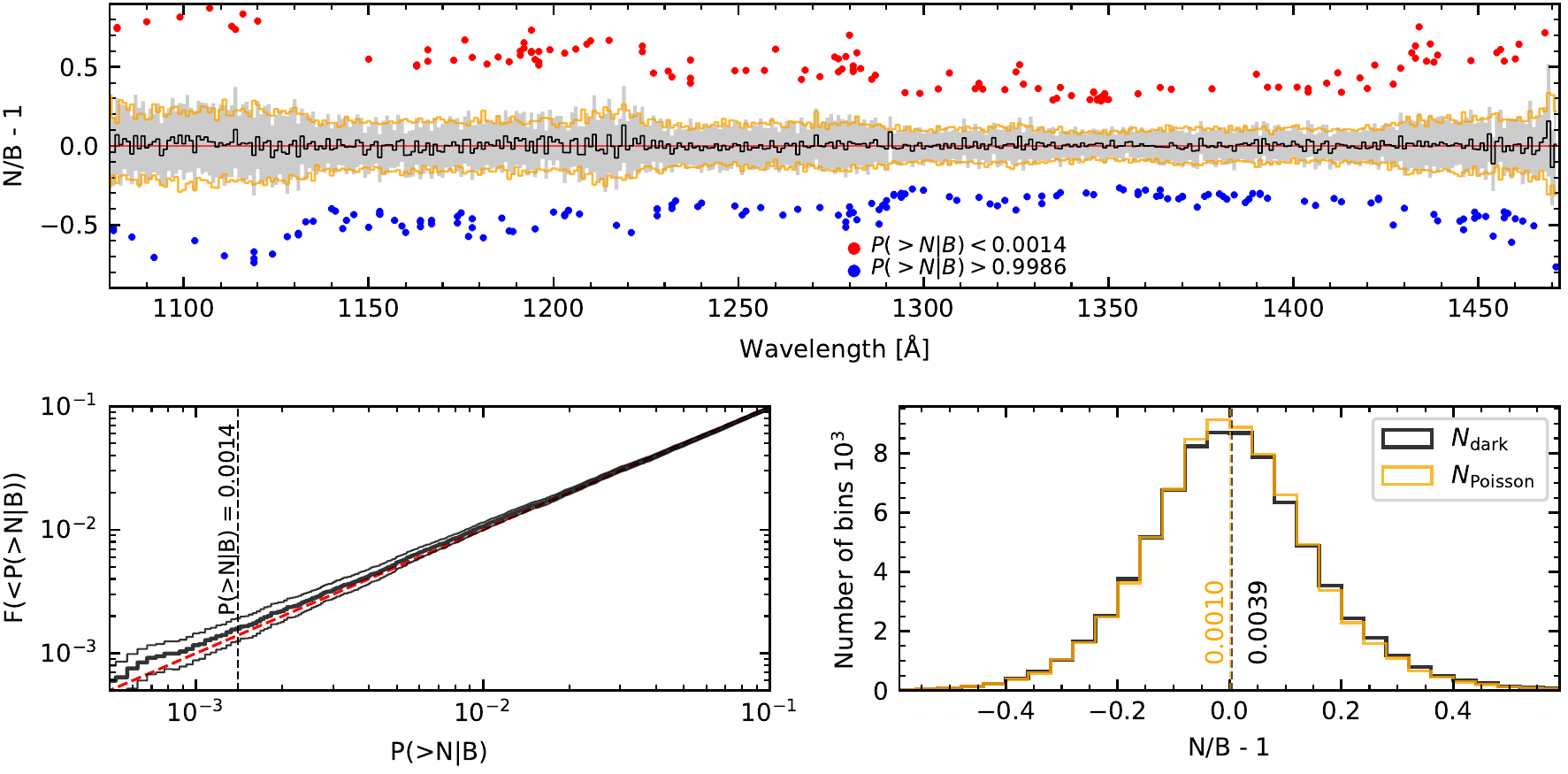}
	\caption{Results of our dark subtraction tests using subsets of darks as data.
	\emph{Upper panel:} Average deviation $N/B-1$ in 1\,\AA\ bins from 200 
	realizations (black) and estimated $1\sigma$ scatter of the 
	distribution (gray). Red and blue points show individual $3\sigma$ 
	deviations with $P(> N|B) \le 0.0014$ and $P(> N|B) \ge 0.9986$, respectively.
    The orange lines indicate the $1\sigma$ scatter of purely Poisson 
    distributed counts $N_\mathrm{Poisson}$ assuming the background model $B$.
    \emph{Lower left panel:} Cumulative fraction of Poisson probabilities 
    per 1\,AA\ bin for measured counts $N$ (thick black) and its $2\sigma$ 
    statistical error (thin black). The dashed red and black lines show 
    identity and $P (> N|B) = 0.0014$, respectively.
    \emph{Lower right panel: } Histogram of deviations $N/B - 1$ in 1\,\AA\ bins 
    for measured counts $N_\mathrm{dark}$ and Poisson distributed counts 
    $N_\mathrm{Poisson}$ according to our modeled dark. The dashed lines show 
    the respective average.}
	\label{fig:bkg_test}
\end{figure*}

\subsection{COS Flat-Fielding and Flux Calibration}

In contrast to \texttt{CALCOS}, \texttt{FaintCOS} preserves Poisson counts in 
co-added spectra, which in turn requires the storage of flat-field and flux 
calibration curves for every sub-exposure. The combined flat-field and deadtime 
correction $K_\mathrm{FF}$ is calculated from the corrected net count rate and 
the gross count rate provided by \texttt{CALCOS} in the extracted sub-exposures.
Individual values are linearly interpolated to provide a continuous flat-field 
curve. The net count rate and the \texttt{CALCOS} flux values yield the 
flux calibration curve.

\subsection{Subtraction of Scattered Geocoronal Ly$\alpha$ Emission}

Scattered geocoronal Ly$\alpha$ emission in G140L spectra is modeled and 
subtracted following \citet{Worseck2016}. The model predicts the corresponding 
counts $B_\mathrm{Ly\alpha}$ and their associated error for data taken at 
the G140L central wavelengths 800\,\AA\ and 1105\,\AA. The 1280\,\AA\ setting 
cannot be corrected, because Ly$\alpha$ emission falls into the detector gap. 
For the G130M grating the sparse geocoronal calibration data sets are 
insufficient to model scattered geocoronal Ly$\alpha$ emission. From our deep 
G130M data we conclude that scattered geocoronal Ly$\alpha$ is 
negligible (Section~\ref{sec:data_reduction}).

\subsection{Co-addition of Spectra Combining Central Wavelength Settings and Data Sets}
\label{sec:co-adding}

Several custom routines have been developed to co-add COS exposures taken at 
several focal-plane offset positions, wavelength settings and/or in 
several visits. \citet{Peeples_cosrep_2017} list common implementations 
\citep{Danforth2010, Keeney2012, Wakker2015, Tumlinson2013}, and describe their 
own approach used for the \textit{HST} Spectroscopic Legacy Archive.
We find that all these routines are not fully applicable because they either 
(1) do not preserve Poisson counts, (2) overestimate statistical errors that 
must be calculated only for the signal part of the Poisson counts 
\citep{FeldmanCousins1998}, or (3) do not properly handle data quality flags.

In \texttt{FaintCOS}, exposures taken at different focal-plane offsets, 
central wavelengths and/or in several visits are co-added in count space on a 
regular wavelength grid. As the resolving power depends on wavelength, 
central wavelength, and the COS Lifetime Position, the bin size can be set 
by the user. In every bin, the pixel exposure time, gross counts, and the 
smoothed total background are summed across all sub-exposures. Only pixels with 
valid science data (data quality flag of zero) contribute to the final co-add. 
Flux calibration and flat-field curves are weighted with the exposure time. 
This is particularly important when combining different central wavelength 
settings that map parts of the wavelength range onto different COS detector segments.

The background error propagation needs to be treated carefully, 
since neighboring pixels of the same exposure are highly covariant due to the 
running average method used for the dark current estimation. 
The covariance is assumed to be negligible for different exposures.
Thus, the total systematic background error is
\begin{equation}
\sigma_B = \sqrt{\sum_{k} \left( \sum_{i=1}^m \sigma_{B}^{ki}\right)^{2}}\quad,
\end{equation}
where exposure $k$ contributes $m$ pixels to the rebinned spectrum.
The flux density per wavelength bin is calculated as
\begin{equation}
\label{eq:flux}
f_\lambda = \frac{(N- B)}{t_{\mathrm{exp}} C K_\mathrm{FF}}\quad,
\end{equation}
with gross counts $N$, 
background counts $B$, 
pixel exposure time $t_{\mathrm{exp}}$, 
flux calibration curve $C$, 
and correction factor $K_\mathrm{FF}$.
The final co-added spectrum is stored in the output table 
\texttt{OBJECT\_spectrum.fits} described in Table~\ref{tab:coadd_result}.

\section{Validation of the Dark Current Model}
\label{sec:dc_validation_test}

We tested our advanced dark subtraction technique by treating dark frames as 
science data. The corrected event lists of individual science exposures 
($t_\mathrm{exp}=$2000--2900\,s) were replaced with those of two randomly 
selected 1330\,s dark exposures taken in a two months period around the 
observation date. The validation data created in this way were run through 
our dark current estimation and co-adding routines.

Figure~\ref{fig:bkg_test} shows the results obtained for 200 realizations of 
the combined data set on HE2QS\,J1630$+$0435 (Table~\ref{tab:obs_data}).
The total exposure time of each validation data set ($42,560$\,s) is comparable 
to the science exposure time ($45,915$\,s). We measured the deviation $N/B - 1$ 
between the measured counts $N$ and estimated dark current model 
$B$ in 1\,\AA\ bins (25 pixels). On average the dark current is slightly 
underestimated by $\lesssim 0.5$\% (Figure~\ref{fig:bkg_test}, lower right panel), 
with localized larger deviations. The small-scale underestimation at 1220\,\AA\ 
is likely due to gain sag around geocoronal Ly$\alpha$ that is not captured by 
our 500-pixel running average. Other deviations are less obvious due to the 
different pixel exposure times, especially at the ends of the spectral range 
and due to the extrapolation of the smoothed dark at the detector edges.
However, the deviations are completely dominated by Poisson noise around the 
estimated dark current. The majority of the bins show a very low average 
deviation of $\le 5$\% which is comparable to the estimated systematic dark 
current error of 1--6\%. The fraction of bins with Poisson probability 
$P \le 0.0014$ ($3\sigma$), which would indicate pixels with statistically 
significant transmission in the scientific data, is slightly higher than 
expected for purely Poisson distributed counts (Figure~\ref{fig:bkg_test}, 
lower left panel). This effect can be explained with our slight underestimation 
of the dark current and its small-scale structure on the detector below 
our 500-pixel averaging scale. 

\section{Decontamination from Geocoronal Emission Lines}
\label{sec:geocor_decont}

\begin{figure}
    \includegraphics[width=\columnwidth]{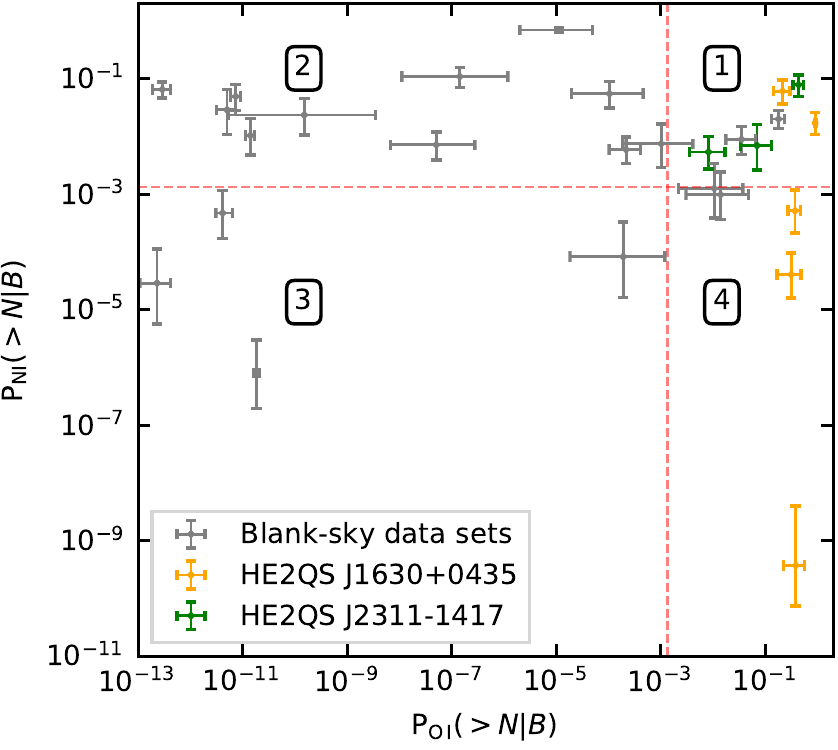}
    \caption{Probability $P$ (Equation~\ref{eq:prob}) to measure $N$ counts in 
    the spectral ranges of geocoronal \ion{N}{1} and \ion{O}{1} during orbital 
    night given the background $B$. Blank-sky data sets and individual science 
    data sets are shown in gray and color, respectively. The red dashed lines 
    indicate a $3\sigma$ detection at $P( > N|B) = 0.0014$. The errors were 
    calculated by varying the background according to the systematic 
    background error.}
    \label{fig:geocoronal_Pvalue}
\end{figure}

To verify the tentatively detected \ion{He}{2} Ly$\beta$ transmission coinciding 
with the geocoronal \ion{N}{1}\,$\lambda$1200 emission line 
(Figure~\ref{fig:day_night_comparison}), we extensively tested the time 
variation of the geocoronal \ion{N}{1}\,$\lambda$1200 and 
\ion{O}{1}\,$\lambda$1304 emission lines. Often it is possible to entirely 
suppress these two emission lines in the affected wavelength ranges by 
considering only part of the data taken during orbital night. We tested the 
correlation of the geocoronal line fluxes during orbital night in archival 
{\it HST}/COS blank-sky 
observations\footnote{https://www.stsci.edu/hst/instrumentation/cos/calibration/airglow}.
We reduced the blank-sky sets covering both \ion{N}{1} and \ion{O}{1} in the 
same way as our science data, restricting the data in the vicinity of the 
lines to night-only data. We calculated the probability $P(>N|B)$ 
(Equation~\ref{eq:prob}) to find at least $N$ counts given the background $B$ 
in the \ion{N}{1} ($1198.5$--$1201.0$\,\AA) and \ion{O}{1} (1301--1307\,\AA) 
regions. Low $P$ values correspond to significant flux.

Figure~\ref{fig:geocoronal_Pvalue} shows the resulting $P_{\mathrm{N\,I}}$ and 
$P_{\mathrm{O\,I}}$ of each blank-sky data set in comparison to the same values 
for individual science data sets. The plot can be divided into four regions for 
the four possible scenarios assuming a $3\sigma$ ($P(>N|B)=0.0014$) detection limit:
(1) no significant flux in both regions,
(2) strong (weak/insignificant) flux in the \ion{O}{1} (\ion{N}{1}) region, 
(3) significant flux in both regions, and 
(4) insignificant flux in the \ion{O}{1} region, but significant flux in 
the \ion{N}{1} region. Scenario 4 is very rare for the blank-sky data sets given 
the $1\sigma$ errors of the $P$ values estimated from the systematic background 
error, so significant \ion{N}{1} emission is generally accompanied by 
strong \ion{O}{1} emission. The HE2QS\,J2311$-$1417 data sets are consistently 
in Scenario 1, showing no residuals in \ion{N}{1} and \ion{O}{1}.
For all but two HE2QS\,J1630$+$0435 data sets it is highly unlikely that the 
residual flux is geocoronal \ion{N}{1}, because \ion{O}{1} is effectively 
eliminated. The two high $P_{\mathrm{N\,I}}$ values are caused by the short 
exposure time of these particular HE2QS\,J1630$+$0435 data sets.
Thus, we conclude that the residual flux at 1200\,\AA\ in the spectrum of 
HE2QS\,J1630$+$0435 is probably \ion{He}{2} Ly$\beta$ transmission 
(Figure~\ref{fig:day_night_comparison}), but due to the fact that all blank-sky 
observations are very short (orbital night $<1200$\,s) we cannot definitely 
determine the origin of the flux. Much longer exposures might show residuals 
comparable to the flux detected in our science data, and we hesitate to draw 
general conclusions from the successful decontamination of 
our HE2QS\,J2311$-$1417 data sets.

\section{Tentative \ion{He}{2} Ly$\beta$ transmission at $z\simeq 3.68$}
\label{sec:likely_lyb_spikes}

\begin{figure}
	\includegraphics[width=\columnwidth]{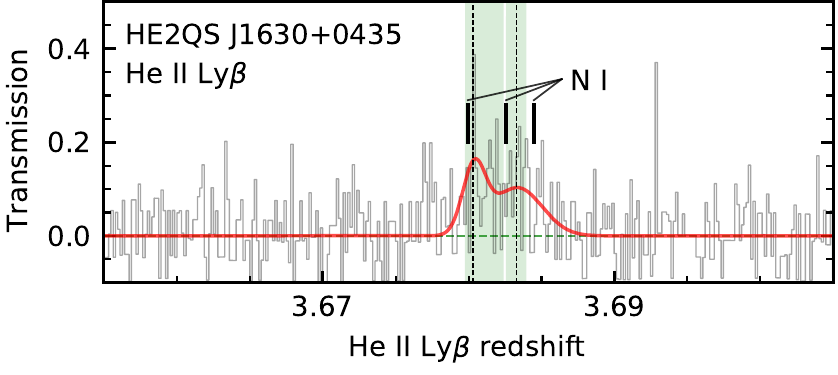}
	\caption{Similar to Figure~\ref{fig:lya_spikes} but for 
	\ion{He}{2} Ly$\beta$ in the spectrum of HE2QS\,J1630$+$0435. 
	Thick vertical lines indicate the position of the geocoronal 
	\ion{N}{1}\,$\lambda$1200 triplet.}
	\label{fig:Lyb_model_2}
\end{figure}

Here we consider the case where the flux at 1200\,\AA\ in the spectrum of 
HE2QS\,J1630$+$0435 is not residual geocoronal \ion{N}{1}, but instead 
intergalactic \ion{He}{2} Ly$\beta$ transmission. Figure~\ref{fig:Lyb_model_2} 
shows its  decomposition into two Gaussian components with fit parameters 
$(z_{\beta,m}, \sigma_m, A_m)$ of $(3.6803, 50\mathrm{\,km\,s^{-1}}, 0.142)$ and
$(3.6833, 109\mathrm{\,km\,s^{-1}}, 0.103)$, respectively.
Following our procedure for the combined posterior in Section~\ref{sec:gamma}, 
we use the occurrence of 
these two \ion{He}{2} Ly$\beta$ and two \ion{He}{2} Ly$\alpha$ spike components to infer a \ion{He}{2} photoionization rate 
$\Gamma_\mathrm{HeII} = (2.5^{+0.7}_{-0.6}) \times 10^{-15}\mathrm{\,s^{-1}}$
at $z = 3.56$--$3.72$ in the HE2QS\,J1630$+$0435 sightline.
This photoionization rate is consistent with our inferences from 
\ion{He}{2} Ly$\alpha$ and \ion{He}{2} Ly$\beta$ at $z \sim 3.6$ at $2\sigma$. 

\clearpage
\bibliography{references}

\end{document}